\documentstyle[aps,psfig]{revtex}
\newcommand{\PP}{{\rm I\hspace{-0.65ex}P}}
\newcommand{\bra}[1]{\mbox{$\langle #1|$}}
\newcommand{\ket}[1]{\mbox{$|#1\rangle$}}
\newcommand{\braket}[2]{\mbox{$\langle #1 | #2 \rangle$}}

\begin{document}
%\preprint{MKPH-T-99-23}

\title
{
%\hfill{\small {\bf MKPH-T-99-23}}\\
{\bf The Role of Meson Retardation in the $NN$ Interaction 
above Pion Threshold}
  \footnote[2]
  {Supported by the Deutsche Forschungsgemeinschaft (SFB 443).}
}
\author
{Michael Schwamb and  Hartmuth Arenh\"ovel}
 \address{ Institut f\"ur Kernphysik,           
  Johannes Gutenberg-Universit\"at,  
  D-55099 Mainz, Germany }
\maketitle

\begin{abstract}
\noindent
A model is developed for the hadronic interaction in the two-nucleon system 
above pion threshold which is based on meson, nucleon and $\Delta$ 
degrees of freedom and which includes full meson retardation in the 
exchange operators. For technical reasons, the model allows 
maximal one meson to be present explicitly. Thus the Hilbert space 
contains besides $NN$ and $N\Delta$ also configurations consisting of 
two nucleons and one meson. For this reason, only two- and three-body 
unitarity is obeyed, and the model is suited for reactions in the two 
nucleon sector, where only one pion is produced or absorbed. Starting from 
a realistic pure nucleonic retarded potential, which had to be renormalized 
because of the additional $\pi$ and $\Delta$ degrees of freedom, a 
reasonable fit to experimental $NN$ scattering data could be achieved.
\end{abstract}


\section{Introduction}\label{introduction}
At present a very interesting topic in the field of medium energy 
physics is devoted to the role of effective degrees of freedom (d.o.f.) in 
hadronic systems in terms of nucleon, meson and isobar d.o.f.\ and their 
connection to the underlying quark-gluon dynamics of QCD. For the study of 
this basic question, the two-nucleon system provides an 
important test laboratory, because it is obviously the simplest nuclear 
system for the study of the nucleon-nucleon interaction in 
$NN$ scattering including deuteron properties, and, furthermore, for 
testing this effective description in other reactions on the deuteron, 
for example, in elastic and inelastic electron scattering, in 
photodisintegration, and meson photo- and electroproduction. 
Moreover, due to the lack of free 
neutron targets, reactions on the deuteron are also an important tool to 
test our present understanding of neutron properties. As an example for 
the latter aspect, we would like to mention the recently much discussed 
determination of the electric form factor $G_{E,n}$ of the neutron in 
$d(\vec{e}, e' \vec{n})p$~\cite{OsH99} and $\vec d(\vec{e}, e'n)p$ 
\cite{PaA99}, or the investigation of the Gerasimov-Drell-Hearn sum rule 
for the deuteron \cite{ArK97}. However, in order to separate the unwanted 
binding effects from the neutron properties, a precise knowledge of the 
structure dependent effects is needed. 
 
Intense efforts over many past decades in experiment and theory have shown 
that for energies below pion threshold a satisfactory description of 
$NN$ scattering and deuteron properties \cite{Mac89,CaS98}, 
as well as photo- and electrodisintegration of the deuteron is achieved 
within the conventional framework of nucleon, meson and isobar d.o.f.
\cite{ArS91,Are99}, although the description of observables is not perfect 
because for certain observables significant discrepancies remain unresolved
as, for example, in elastic electron deuteron scattering \cite{ArR99}. At 
higher energies, above pion threshold our theoretical understanding of the 
various experimental data is much less well settled. Even for deuteron 
photodisintegration (for a detailed review see \cite{ArS91}), none of the 
various theoretical approaches like the diagrammatic method of Laget 
\cite{Lag78}, the framework of nuclear isobar configurations in the impulse 
approximation \cite{Sch95}, the unitary three-body model \cite{TaO89} or the 
coupled channel approach (CC) \cite{LeA87,WiA93} is able to describe in a 
satisfactory manner the whole set of experimental data on differential 
cross sections and polarization observables for energies covering the whole 
$\Delta$ resonance region. 

A common feature of most of these approaches is the extensive use of the 
static limit for the meson propagator which enters the hadronic interaction 
and the electromagnetic two-body exchange current operators, although there 
is little justification for that in view of the relatively high excitation 
energies involved. Also from the point of view of special relativity, 
a non-instantaneous interaction would be required. 
The reason, why this static approximation is still used, is the enormous 
simplification of the operator structure becoming local, and thus is much 
simpler to evaluate numerically. It has already been conjectured in 
\cite{WiA93}, that one main reason for the above mentioned failure of the 
theoretical description lies in the neglect of meson retardation in the 
meson exchange operators. Indeed, first results \cite{Sc198,Sc298} based on 
the thesis of M.\ Schwamb~\cite{Sch99} have shown that 
a much better description of deuteron photodisintegration compared to the 
static approaches is obtained if meson retardation is retained. 

In this paper, we want to present a realistic, but still
tractable model for a retarded hadronic interaction within a nonrelativistic 
framework which is suitable to be used as input in electromagnetic reactions 
on the deuteron like, for example, photo- and electrodisintegration or 
pion production. There exists already a variety of models for treating 
retardation in $NN$ scattering based on three-body theory, for example, the 
work of Kloet and Silbar~\cite{KlS80a} and Tanabe and Ohta~\cite{TaO88}. In these 
calculations, however, the driving force is
basically one-pion-exchange so that the lower partial waves in $NN$ scattering 
cannot be described reasonably well, even at energies below pion threshold. 
The neglected short-range interaction is, however, of crucial importance for 
electromagnetic and hadronic breakup reactions on the deuteron. Tanabe and 
Ohta, for example, have "patched up" this problem in photodisintegration of 
the deuteron~\cite{TaO89} by mixing retarded and static 
frameworks, namely the static Paris potential for the final state 
interaction with respect to the nucleon one-body current and the $\pi$ MEC, 
whereas for the $\Delta$ contributions they employ a three-body model in which
heavy meson exchange has been neglected. Kloet and 
Silbar on the other hand have taken a retarded pion exchange and static heavy 
meson exchange in an improved model of $NN$ scattering~\cite{KlS80b}. 
Also other treatments of retardation in the $NN$-interaction like, e.g.\ 
Refs.~\cite{ElH88,BuS92} have some drawbacks as will be discussed later. 
The shortcomings of these approaches have been overcome to a large extent in 
the present model which ensures, due to the use of the 
Elster-Potential~\cite{Els86,ElF88} as an 
input, a reasonable description of all relevant partial waves below and above 
pion threshold within a consistent framework in time-ordered 
perturbation  theory. 

Thus we will present here the basic framework of a model for the 
hadronic interaction in which retardation in the exchange operators is 
fully included. We would like to emphasize the fact that this formalism 
can be applied to any reaction on the two-nucleon system for excitation 
energies up to about 500 MeV in which not more than one pion is produced 
or absorbed. But in this paper we will restrict ourselves to the hadronic 
interaction in studying $NN$ scattering and deuteron properties 
only in order to fix all free hadronic parameters of 
our model. The application of this formalism to other hadronic reactions 
like $\pi d$ scattering as well as to electromagnetic reactions will be 
deferred to forthcoming papers. 

The conceptual basis and the main features 
of our model are laid out in Sect.~\ref{kap_hadonische_ww}. In view 
of the fact that conventional retarded interactions in pure nucleonic space 
are energy dependent and thus non-hermitean, we enlarge the Hilbert space 
by considering explicitly meson and $\Delta$ d.o.f.\ in order to start 
with a hermitean hamiltonian. The retarded interaction 
is then generated by meson-nucleon  and $\pi N \Delta$ vertices. For reasons 
of simplicity, we restrict ourselves in this work to configurations 
with only one meson present besides the baryons. Special attention is laid on 
the question of nucleon dressing and the corresponding renormalization of 
operators as well as the fulfilment of two- and three-body unitarity. With 
respect to the latter, we have incorporated the $\pi d$ channel in a 
realistic model showing some interesting effects. The $NN$ scattering 
$T$ matrix is derived, and the structure of the deuteron is discussed. 
A field theoretical realization in the form of a one-boson-exchange model 
is developed in Sect.~\ref{kap3_kap_ham3} with inclusion of the $\Delta$ 
isobar which is conceptually similar to the work of the Bonn 
group~\cite{MaH87} but differs in some essential details. Then we will 
present and discuss in Sect.~\ref{kap6} the results for $NN$ scattering 
comparing them with experiment as well as with other theoretical approaches. 
Finally, Sect.~\ref{summary} contains a summary and an outlook.

\section{Basic considerations}\label{kap_hadonische_ww}

\subsection{The Hilbert space}\label{kap_hilbert}
Our model for the description of a nuclear system allows besides $n$
nucleons configurations with one additional meson present or  
where one nucleon is replaced by a $\Delta$ isobar. It 
is similar (but not identical) to the approach of Sauer and collaborators 
\cite{BuS92,Sau86,PoS87,PeG92} which is also used in \cite{WiA93,WiA95,WiA96}.
Thus the model Hilbert space ${\cal H}^{[n]}$ is subdivided into three 
orthogonal spaces according to the different configurations containing 
either $n$ bare nucleons $({\cal H}^{[n]}_{\bar N })$, $n-1$ nucleons and 
one $\Delta$ (${\cal H}^{[n]}_{\Delta})$, or $n$ nucleons and one 
meson $({\cal H}^{[n]}_{X})$, i.e.\
\begin{equation}\label{kap3_hilbert}
{\cal H}^{[n]} = {\cal H}^{[n]}_{\bar N } \oplus
{\cal H}^{[n]}_{\Delta} \oplus 
{\cal H}^{[n]}_{X}\,.
\end{equation}
The ``bar'' indicates a bare nucleon to be distinguished from the 
corresponding physical nucleon, denoted without a bar. This distinction 
is necessary in order to take into account the dressing of a bare nucleon by
meson-nucleon loops. For the $\Delta$ isobar we will not make this 
distinction, because the self energy 
contributions from $\pi N$ loops will be retained explicitly, whereas 
the dressing of the bare nucleon to become a physical nucleon will be 
incorporated into the effective operators by dressing factors. In 
${\cal H}^{[n]}_{X}$ only one meson is present besides 
$n$ nucleons, i.e., no components with two or more mesons are taken into 
account (one-meson-approximation). This limitation creates some pathologies
as will be pointed out later. In detail it means 
\begin{equation}\label{kap3_qaufsp}
{\cal H}^{[n]}_{X} = \bigoplus 
\limits_{x \in \{\pi,\, \eta,\, \sigma,\, \delta,\, \omega,\, \rho\}}
{\cal H}^{[n]}_{x}\,,
\end{equation}
considering as mesons $\pi$, $\eta$, $\sigma$, $\delta$, $\omega$, and $\rho$.

In order to distinguish the various sectors of ${\cal H}^{[n]}$, we introduce 
corresponding projection operators by
\begin{equation}\label{kap3_projektoren_delta}
P_{{\bar N}}  {\cal H}^{[n]}  =
{\cal H}^{[n]}_{\bar N}\,, \qquad 
P_{\Delta}{\cal H}^{[n]}
={\cal H}^{[n]}_{\Delta}\,,
\qquad  P_{X}{\cal H}^{[n]} 
={\cal H}^{[n]}_{X}\,,\mbox{ and }P=P_{\bar N}+P_{\Delta}\,. 
\end{equation}
The latter is introduced since later we also will need the projection 
operator on the combined pure baryon subspace 
${\cal H}^{[n]}_{\bar N}\oplus {\cal H}^{[n]}_{\Delta}$. 
Moreover, in view of the different subspaces in the meson-nucleon sector, 
it is useful to decompose the projection operator $P_{X}$ into a sum of 
six orthogonal projectors corresponding to the six mesons considered
\begin{equation}\label{kap3_q}
P_{X} = \sum_{x \in \{\pi,\, \eta,\, \sigma,\, \delta,\, \omega,\, \rho\}}
P_{x}\, \mbox{ with }  
P_{x}  {\cal H}^{[n]} ={\cal H}^{[n]}_{x} \, \mbox{ for } 
 x \in \{ \pi, \eta, \sigma, \delta, \omega, \rho \}\, .\label{projektoren3}
\end{equation}
Using the notation 
\begin{equation}\label{matrix_kurz}
\Omega_{\alpha \beta} = P_{\alpha} \Omega P_{\beta}\, ,
\qquad \alpha,\,\beta \in \{\bar N,\, \Delta,\, X\}\,,
\end{equation}
any operator $\Omega$ acting in  ${\cal H}^{[n]}$ can be written as a 
symbolic $3\times 3$ matrix 
\begin{equation}\label{matrixnotation}
\Omega = \left( \begin{array}{ccc}
\Omega_{{\bar N}{\bar N}} & 
\Omega_{{\bar N}\Delta} & 
\Omega_{{\bar N} X} \\
\Omega_{\Delta {\bar N}} & 
\Omega_{\Delta \Delta} & 
\Omega_{\Delta X} \\
\Omega_{X {\bar N}} & \Omega_{X \Delta} & \Omega_{X X} 
                \end{array} \right) .
\end{equation}

\subsection{The Hamiltonian}\label{kap_ham_1}
The Hamilton operator $H$  of the model can be divided into a diagonal 
kinetic part and an interaction describing the emission 
and absorption of a meson by a baryon and, in addition, an ab initio 
baryon-baryon interaction
\begin{equation}\label{kap3_h1}
H = \bar H_0 + H_I\,,
\end{equation}
where the bar indicates that it refers to bare baryons. In particular, the
nucleon kinetic energies are determined by the bare nucleon mass $M_{\bar N}$.
However, due to the truncation of the Hilbert space with respect to the 
number of mesons, no explicit dressing of the bare nucleons is possible in 
${\cal H}^{[n]}_{\Delta} \oplus {\cal H}^{[n]}_{X}$. This is one of the 
pathologies which one encounters in the one-meson-approximation. 
Therefore, we will use the physical instead of the bare nucleon mass in 
${\cal H}^{[n]}_{\Delta} \oplus {\cal H}^{[n]}_{X}$, 
so that the formally suppressed meson-nucleon loops can be taken into account
at least effectively by the physical nucleon mass. For simplicity, we 
use in ${\cal H}^{[n]}_{\Delta}$ the nonrelativistic expressions for
the kinetic energies of nucleon and $\Delta$ isobar because we will treat the 
$\Delta$ nonrelativistically. This approximation is, however, not crucial.
For later purposes we will introduce as kinetic energy in addition a second 
type of one-body operator $H_0$ which differs from $\bar H_0$ 
in the pure nucleonic sector only, referring to physical nucleon 
kinetic energies also in ${\cal H}^{[n]}_{\bar N}$. 
Thus we have in detail
\begin{eqnarray}
\bar H_{0,\,{\bar N}{\bar N}} &=& \sum_{j=1}^n {h}_{\bar N}(j)\,, 
\label{h00_1}\\
H_{0,\,{\bar N}{\bar N}} &=& \sum_{j=1}^n {h}_{N}(j)\,, 
\label{kap3_h0nn1}\\
\bar H_{0,\,\Delta \Delta} &=& H_{0,\,\Delta \Delta} =\sum_{j=1}^{n-1}
{h}^{nr}_{N}(j) +{h}_{\Delta}\,,
\label{h00_2} \\
\bar H_{0,\,XX} &=& H_{0,\,XX}=\sum_{j=1}^n {h}_{N}(j) + 
{h}_{X}\,,
\label{h00_3}
\end{eqnarray}
with the various kinetic energies  (with masses $m_x,
x \in \{ \pi,\, \eta,\, \sigma,\, \delta,\, \omega,\, \rho \}$,
$M^0_{\Delta}$,  $M_{\bar N}$ and $M_N$, respectively)
\begin{eqnarray}
\bra{{\bar N}(\vec{p}^{\, \prime})}{h}_{\bar N}\ket{{\bar N}(\vec{p}\,)}
=  e_{\bar N}(p)\,\delta\left(\vec{p}^{\, \prime} - \vec{p}\, \right) 
\,, &\qquad& e_{\bar N}(p) = \sqrt{{M^2_{\bar N}} + p^2} \,, 
\label{zus_ein-meson1}\\
\bra{{\bar N}(\vec{p}^{\, \prime})}{h}_{N}\ket{{\bar N}(\vec{p}\,)}
= e_{N}(p) \,\delta\left(\vec{p}^{\, \prime} - \vec{p}\, \right)
\, , &\qquad& e_N(p) = \sqrt{ M^2_N + p^2} \,,   \label{zus_ein-meson2}\\
\bra{{\bar N}(\vec{p}^{\, \prime})}{h}^{nr}_{N}\ket{{\bar N}(\vec{p}\,)}
= e^{nr}_{N}(p) \,\delta\left(\vec{p}^{\, \prime} - \vec{p}\, \right) 
\, , &\qquad& e^{nr}_N(p) = M_N + \frac{p^2}{2 M_N}\,,  \label{zus_ein-meson3}\\
\bra{\Delta(\vec{p}^{\, \prime})}{h}_{\Delta} 
\ket{\Delta(\vec{p}\,)}
= e_{\Delta}(p)\,\delta\left(\vec{p}^{\, \prime} - \vec{p}\, \right) 
\,, &\qquad& e_{\Delta}(p) = M^0_{\Delta} + 
\frac{ p^2}{2 M^0_{\Delta}}  \,,\label{zus_ein-meson4}\\
\bra{x(\vec{q}^{\, \prime})}{h}_{x} \ket{x(\vec{q}\,)}
= (2 \pi)^3 \,2 \,\omega_x(q)\,
\delta\left(\vec{q}^{\, \prime} - \vec{q} \, \right)\,,
&\qquad& \omega_{x}(p) = \sqrt{m_{x}^{2} + p^2} \,, 
 \quad \mbox{for}  \,\, x \in \left\{\pi, \eta, \sigma, 
\delta, \omega, \rho \right\}\,.
\label{zus_ein-meson5}
\end{eqnarray}

\begin{figure}[hp]
\centerline{\psfig{figure=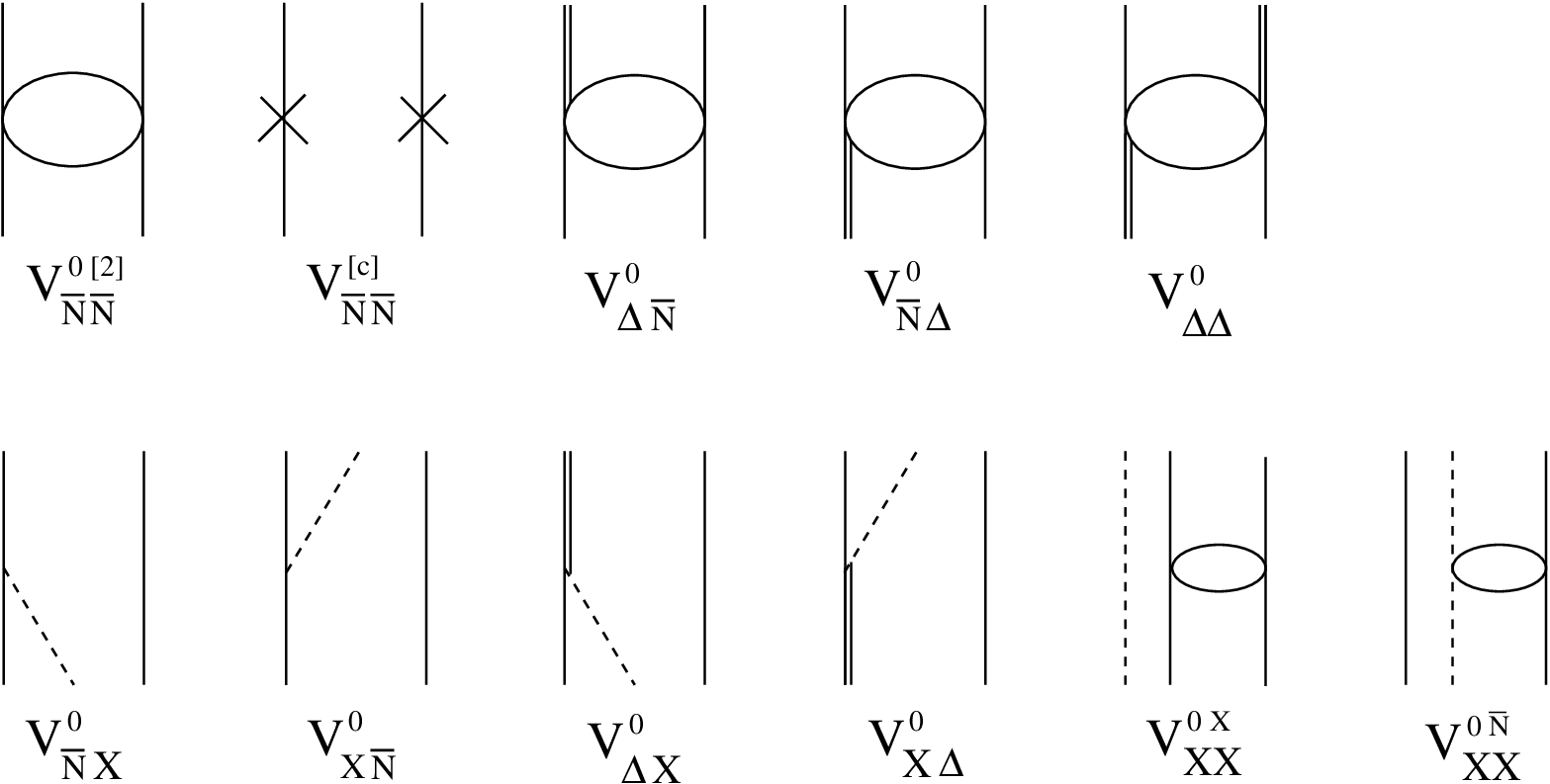,width=11cm,angle=0}}
\vspace{0.5cm}
\caption{Diagrammatic representation of the various components of 
 $V^0$. The open ellipse symbolizes a given hermitean two-body interaction.
 The one-nucleon counter term $v^{[c]}$ is indicated by a cross.}
\label{potentialuebersicht}
\end{figure}

In view of the two choices for the kinetic energy, one obtains besides 
the interaction $H_I$, defined in (\ref{kap3_h1}) another interaction 
operator $V^0$ as defined via 
\begin{equation}\label{kap3_h2}
H = H_0 + V^0 \,.
\end{equation}
The various components of the interaction $V^0$ are depicted in 
Fig.~\ref{potentialuebersicht}. First of all, $V^0$ contains, because of
\begin{equation}
V^0 = H_I + \bar H_0 - H_0\,,
\end{equation}
besides the interaction $H_I$, a diagonal counter term 
\begin{equation}\label{kap3_vc}
V^{[c]} = \bar H_0 - H_0 \,,
\end{equation} 
which is nonzero in ${\cal H}^{[n]}_{\bar N}$ only, and which
is a pure one-body operator 
\begin{equation}\label{kap3_counter_zerlegung}
V^{[c]}_{\bar N\bar N} = \sum_{j=1}^n v^{[c]}_{\bar N\bar N}(j) \,,
\mbox{ with }
v^{[c]}_{\bar N\bar N} = {h}_{\bar N} - {h}_{N}\,.
\end{equation}
As already mentioned, we allow in addition in $V^0_{PP}$ a two-body part 
$V^{0\,[2]}_{PP}$ which describes an ab initio given hermitean interaction 
between two baryons, i.e.,
\begin{equation}\label{kap3_vnn}
V^0_{PP} = V^{[c]}_{PP} + V^{0\,[2]}_{PP}\,,
\end{equation} 
which will be specified later. The nondiagonal components $V^0_{X {\bar N}}$ 
and   $V^0_{X \Delta}$ are one-body operators 
\begin{eqnarray}
V^0_{X{\bar N}} &=& \sum_{j=1}^n v^0_{X{\bar N}}(j) \,,\quad
V^0_{X \Delta} = \sum_{j=1}^n  v^0_{X \Delta}(j) \, , \label{kap3_vdn2}
\end{eqnarray}
describing the emission of a meson by a baryon. 
The remaining interaction $V^0_{XX}$ consists in the one-meson-approximation 
in principle of two parts
\begin{equation}\label{kap3_vqq1}
V^0_{XX} = V^{0\,\bar N}_{XX} + 
\sum_{x\in\{ \pi,\, \eta,\, \sigma,\, \delta,\, \omega,\, \rho\}}
V^{0\,x}_{XX} \, .
\end{equation}
The first one describes a meson-nucleon interaction with the other nucleons 
as spectators and the second one two interacting nucleons with a meson $x$ 
and the remaining nucleons as spectators (see Fig.~\ref{potentialuebersicht}).
In consequence, $V^{0\,{\bar N}}_{XX}$ is a one-body operator
\begin{equation}\label{kap3_vqqx}
V^{0\,{\bar N}}_{XX} = \sum_{i=1,2} v^0_{XX}(i) \, .
\end{equation}

In the previous work of Wilhelm {\it et al}.\ \cite{WiA93,WiA95,WiA96} and 
Bulla {\it et al}.~\cite{BuS92}, $V^0_{XX}$ was neglected completely for 
practical reasons. Above the $\pi d$ threshold, however, this approximation 
leads to several problems, in particular for three-body unitarity 
(see the  discussion in Sect.~\ref{kap3_vqq_uni}),
which are solely due to the neglect of $V^{0\,\pi}_{XX}$. We therefore set 
\begin{equation}
V^{0\,{\bar N}}_{XX}  \equiv 0 \,, \label{kap3_vqq}
\end{equation}
and for $x \neq \pi$
\begin{equation}
V^{0\,x}_{XX} \equiv 0 \,,\label{kap3_vqq_x}
\end{equation}
whereas only $V^{0\,\pi}_{XX}$ will be retained nonzero.

\subsection{$NN$ scattering}\label{kap_scatter}

Now we will consider $NN$ scattering in the two-nucleon sector. The 
renormalization of a free nucleon state $\ket{N;\,\vec{p}\,}$ is sketched 
briefly in Appendix~A. The scattering states of two physical 
nucleons $\ket{NN;\,\vec{p}, \alpha}^{(\pm)}$ 
in the c.m.\ system with the asymptotic free relative momentum $\vec{p}$, 
energy $E^{NN}_{p}=2 e_N(p)$ and a complete set of internal 
quantum numbers  $\alpha$ are given by
\begin{equation}\label{kap3_nn_s_mod}
\ket{NN;\,\vec{p}, \alpha\,}^{(\pm)}
= N_{[2]}^{-1}(\vec{p}\,) \Big( 1+
G_{0}(E^{NN}_{p} \pm i\epsilon)  
T^0( E^{NN}_{p} \pm i\epsilon) 
\Big) \ket{\bar N \bar N;\,\vec{p}, \alpha}\, ,
\end{equation}
where $N_{[2]}(\vec{p}\,)$ is a renormalization constant which appears 
because both, the bare as well as the physical free two-nucleon states are 
normalized to the $\delta$ function. The transition amplitude $T^0$ 
satisfies the Lippmann-Schwinger equation
\begin{eqnarray}
T^0(z) &=& V^0  +  V^0   G_0(z) T^0(z)\,\, \label{kap3_u_mod} 
\end{eqnarray}
with the free propagator $G_0$ 
\begin{equation}\label{kap3_g0_2}
G_0(z) =(z-H_0)^{-1}\, ,
\end{equation}
which is represented in Fig.~\ref{fig_G_0}. As is shown in detail in 
the Appendix~B, one finds for the $T$ matrix element 
\begin{eqnarray}
\bra{NN;\,\vec{p}^{\,\prime}, \alpha'} T(E^{NN}_{p}+i\epsilon)
 \ket{NN;\,\vec{p}, \alpha}  & =& N_{[2]}^{-2}(p) 
\bra{{\bar N}{\bar N};\,\vec{p}^{\,\prime}, \alpha'}
 T^0(E^{NN}_{p}+i\epsilon)
 \ket{{\bar N}{\bar N};\,\vec{p}, \alpha} \nonumber\\ 
&=& \bra{{\bar N}{\bar N};\,\vec{p}^{\,\prime}, \alpha'}
 T^{con}_{{\bar N}
 {\bar N}}(E^{NN}_{p}+i\epsilon))
 \ket{{\bar N}{\bar N};\,\vec{p}, \alpha}\, .
 \label{kap3_unn2_mod}
\end{eqnarray} 
Here, $T^{con}$ - the superscript ``$con$'' refers to connected diagrams 
(see Appendix~B) - obeys a Lippmann-Schwinger equation
\begin{equation}\label{kap3_tpp3_mod}
 T^{con}_{PP}(z) =  {\widehat R}(z)\, V^{con}_{PP}(z)\,{\widehat R}(z)
 + {\widehat R}(z)\, V^{con}_{PP}(z) \,{\widehat R}(z)\,
 G_0(z) \,T^{con}_{PP}(z)\,. 
\end{equation}
Its driving term contains a ``renormalized'' interaction  
\begin{eqnarray}
V^{con}_{PP}(z) &=&({\widehat Z}_{[2]}^{os})^{-1}\,V^{0,\,con}_{PP}(z)\,
 ({\widehat Z}_{[2]}^{os})^{-1}\,,
\label{kap3_renorm_mat2_5_mod}
\end{eqnarray}
where $V^{0,\,con}_{PP}$, as defined in Appendix B, comprises besides $\pi N$ 
loop contributions to the $\Delta$ self energy the genuine retarded baryon-baryon 
interaction. Its diagrammatic representation is shown in Fig.~\ref{fig_V_0_con}. 
Furthermore, the ``dressing operator'' is given by
\begin{equation}\label{kap3_renorm_mat_r1}
{\widehat R}(z) = {\widehat Z}_{[2]}^{os}\,
{\widehat Z}_{[2]}^{-1}(z)\,.
\end{equation}
Here the two-body 
renormalization operator ${\widehat Z}_{[2]}$ is defined as
\begin{eqnarray}
{\widehat Z}^2_{[2]}(z) &=&  1 + \int dz'\,\delta(z'-H_0)\,
\left[V^0_{{\bar N}X}\,G_0(z')\,G_0(z)\,V^0_{X{\bar N}}
\right]_{dis}\,, 
\label{kap3_renorm_mat_z2_2}
\end{eqnarray}
where the subscript ``$dis$'' refers to disconnected diagrams 
(see Appendix~B), and which differs from unity in 
${\cal H}^{[2]}_{\bar N }$ only. Its onshell value is 
\begin{eqnarray}
({\widehat Z}^{os}_{[2]})^2 &=&  1 + \int dz'\,\delta(z'-H_0)\,
\left[V^0_{{\bar N}X}\,G_0(z')\,G_0(z')\,V^0_{X{\bar N}}
\right]_{dis}\,. 
\label{kap3_renorm_mat_z2_os}
\end{eqnarray}
Thus $NN$ scattering is unambiguously fixed by the onshell matrix 
element of $T^{con}_{{\bar N}{\bar N}}$. 

\begin{figure}[hp]
\centerline{\psfig{figure=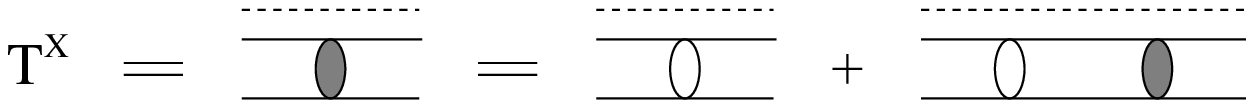,width=12cm,angle=0}}
\vspace{0.5cm}
\caption{Diagrammatic representation of the $NN$ scattering matrix 
$T^X(z)$ in the presence of a spectator meson.}
\label{fig_t_X}
\end{figure}

For later applications we need also the offshell form of $T^{con}_{PP}(z)$ 
for which one finds the following expression  
\begin{equation}\label{kap3_tpp_cal1}
 T^{con}_{PP}(z) = G_0^{-1}(z)
 \left\{ G_0^{(\Delta)}(z)+G_0^{(\Delta)}(z) {\widetilde T}^{con}_{PP}(z) 
 G_0^{(\Delta)}(z) \right\} G_0^{-1}(z)-G_0^{-1}(z)\,.
\end{equation}
The auxiliary amplitude ${\widetilde T}^{con}_{PP}(z)$ is given by the
integral equation 
\begin{equation}
 {\widetilde T}^{con}_{PP}(z) = V^{con}_{[2]\,PP}(z) + 
 V^{con}_{[2]\,PP}(z)  G_0^{(\Delta)}(z){\widetilde T}^{con}_{PP}(z) \,, 
 \label{kap3_tpp_cal2}
\end{equation}
where the driving terms $V^{con}_{[i]\,PP}(z)$ ($i=1,2$) are 
defined by separating in $V^{con}_{PP}(z)$ the $\pi N$ loop contributions 
to the $\Delta$ self energy, subdividing it into two parts
\begin{equation}\label{kap3_z1}
V^{con}_{PP}(z) = V^{con}_{[1]\,PP}(z) +  V^{con}_{[2]\,PP}(z)\,,
\end{equation}
where
\begin{eqnarray}
V^{con}_{[1]\,PP}(z) &=& \left[ V_{\Delta  X} G_0(z)
  V_{X \Delta} \right]_{dis}  \label{kap3_z2}\,, \\
V^{con}_{[2]\,PP}(z) &=&
V^{[2]}_{PP} + 
 \left[V_{PX} G_0(z) V_{XP} \right]_{con}
 + V_{PX} G_0(z)T^X(z)G_0(z) V_{XP} \label{kap3_z3}\,,
\end{eqnarray}
with the renormalized interactions
\begin{eqnarray}
 V^{[2]}_{PP}
&=&({\widehat Z}_{[2]}^{os})^{-1}\,V_{PP}^{0\,[2]} \,
({\widehat Z}_{[2]}^{os})^{-1}\,, \label{renorm1}\\
V_{XP}&=&({\widehat Z}_{[2]}^{os})^{-1}\,V_{XP}^0\,
({\widehat Z}_{[2]}^{os})^{-1}\,. \label{renorm2}
\end{eqnarray}
Here, $T^X(z)$ is the $NN$ scattering matrix in the presence of a spectator 
meson fulfilling 
\begin{equation}\label{kap3_tX_mod}
T^X(z) = V^0_{XX} + V^0_{XX}\,G_0(z)\,T^X(z)\,.
\end{equation}
Its diagrammatic representation is shown in Fig.~\ref{fig_t_X}.
Thus the term $V^{con}_{[1]\,PP}(z)$ contains solely the intermediate 
$\pi N$ loop contributions to the $\Delta$ propagator. Furthermore, we have 
introduced in (\ref{kap3_tpp_cal1}) a ``dressed'' propagator
\begin{equation}
 G_0^{(\Delta)}(z) = G_0(z)+G_0(z) V^{con}_{[1]\,PP}(z) G_0^{(\Delta)}(z) \,,
 \label{kap3_g0d}
\end{equation}
which takes into account the dressing of the $\Delta$ in 
${\cal H}^{[2]}_{\Delta}$. In ${\cal H}^{[2]}_{\bar N }$ and 
${\cal H}^{[2]}_{X}$, $G_0^{(\Delta)}(z)$ is identical to the 
free propagator, whereas in ${\cal H}^{[2]}_{\Delta}$ one gets a simple 
connection between $G_0^{(\Delta)}(z)$ and the $\Delta$ propagator 
$g_{\Delta}(z)$ in the one-$\Delta$ sector (see Sect.~\ref{kap3_ndeltavertex})
\begin{equation}
\bra{{\bar N}\Delta;\,\vec{p}^{\, \prime}}
 G_{0\Delta \Delta}^{(\Delta)}(z) 
 \ket{{\bar N} \Delta;\,\vec{p}\,} 
= \bra{\Delta;\,\vec{p}^{\, \prime}}
g_{\Delta} \left( z-M_N- \frac{\vec{p}^{\,2}}{2\mu_{N\Delta}} \right)
\ket{\Delta;\,\vec{p}\,}\,,
 \label{kap3_gd03}
\end{equation}
where $\vec{p}$ is the relative momentum of 
the $\bar N\Delta$ system, and its reduced mass is denoted by 
\begin{equation}\label{kap3_mu1}
 \mu_{N\Delta}= \frac{M^0_{\Delta} M_N}{M^0_{\Delta}+M_N}\,.
\end{equation}

In addition, we need the baryonic {\it and} mesonic components 
of the $NN$ scattering states for which one finds 
\begin{eqnarray}
P_{\bar N} \ket{NN;\,\vec{p},\, \alpha}^{(\pm)}
 &=& 
   \frac{{\widehat R}(z)}{
{\widehat Z}^{os}_{[2]}} \left( 1+ G_0(z)
 {\widetilde T}^{con}_{\bar N\bar N}(z) \right)
 \ket{\bar N\bar N;\,\vec{p},\, \alpha}\,,
\label{kap3_nn_s2_mod_mod_1}\\
  P_{\Delta} \ket{NN;\,\vec{p},\, \alpha}^{(\pm)} &=&
  G_0^{(\Delta)}(z){\widetilde T}^{con}_{\Delta \bar N}(z) 
\ket{\bar N\bar N;\,\vec{p},\, \alpha}\,,\\
\label{kap3_nn_s2_mod_mod_2}
P_X\ket{NN;\,\vec{p},\, \alpha}^{(\pm)}
 &=&  G^X(z)\Big( V_{X\bar N}\,{\widehat Z}^{os}_{[2]} +
  V_{X\Delta}\Big)
 \ket{ N N;\,\vec{p},\, \alpha} \,,
\label{kap3_nn_s2_2_mod_a}
\end{eqnarray}
where $z= E^{NN}_{p} \pm i \epsilon$, 
and $G^X(z)$ describes the propagation of two interacting nucleons in the 
presence of a spectator meson  
\begin{eqnarray}
 G^X(z)  & =&  (z-H_{0,\,XX}-V^0_{XX})^{-1} \nonumber\\
 & =& G_0(z) + G_0(z)\,T^X(z)\,G_0(z)\,.
 \label{kap3_vqq3_mod}
\end{eqnarray}

\subsection{The deuteron}\label{kap3_deuteron}
Due to its vanishing isospin, the deuteron cannot contain a 
$\bar N \Delta$ component. Thus, the deuteron state $\ket{\bar d}$ can be 
separated into a nucleonic and a mesonic component 
\begin{equation}\label{kap3_deut1}
\ket{\bar d} =  P_{\bar N}  \ket{\bar d} + P_X \ket{\bar d}\,,
\end{equation}
which can be determined by the Schr\"odinger equation in the c.m.\ frame
\begin{equation}\label{kap3_deut2}
\left( H_0 + V^0 \right) \ket{\bar d} = M_d \ket{\bar d}\,\, 
\end{equation}
with $M_d = 2 M_N - \varepsilon_B$ as deuteron mass and $\varepsilon_B$ 
its binding energy. Eliminating the mesonic component, we introduce 
the purely nucleonic part $P_{\bar N}  \ket{\bar d}$ as an effective 
renormalized deuteron state by 
\begin{equation}\label{kap3_dstrich}
\ket{d} = {\widehat Z}_{[2]}(M_d) P_{\bar N} \ket{\bar d}\, ,
\end{equation}
with the normalization $\braket{d}{d} = 1$. It is straightforward to show 
that $\ket{d}$ obeys the equation
\begin{equation}\label{kap3_zwischen3}
\left( H_0 +  {\widehat R}(M_d) \Big(V^{[2]}_{\bar N \bar N} +
\left[V_{\bar N X} G_0(M_d) V_{X\bar N }  \right]_{con} \Big)
{\widehat R}(M_d)  \right) \ket{d} = M_d \ket{d}\,,
\end{equation}
which contains only renormalized quantities. From the effective state 
$\ket{d}$ the original deuteron state $\ket{\bar d}$ is obtained by
\begin{equation}\label{kap3_deuteron_final}
\ket{\bar d} = \frac{1}{N_d} \left(\frac{{\widehat R}(M_d)}
{{\widehat Z}^{os}_{[2]}}
+ G_0(M_d) V_{X\bar N } {\widehat R}(M_d) \right) \ket{d}\,,
\end{equation}
with the renormalization constant $N_d$  
\begin{eqnarray}
N_d^2 &=& \bra{d} \left(
1 - \frac{\partial}{\partial z} \left.\left\{
{\widehat R}(z)    \left[ V_{\bar N X} G_0(z) 
 V_{X\bar N } \right]_{con}  {\widehat R}(z) \right\}  \right|_{z=M_d}
\right) \ket{d} \, .
 \label{kap3_deuteron_norm1} 
\end{eqnarray}

Due to the absence of $\bar N\Delta$ components and the vanishing 
of $V^0_{XX}$ in $(t=0)$ channels, the quantity $V^{[2]}_{\bar N \bar N }$  in 
(\ref{kap3_zwischen3}) can be put equal to zero in retarded calculations. 
On the other hand, in static approaches, because of the choice 
$V^0_{\bar N X}=0$, one has to identify $V^{0\,\,[2]}_{NN}$ with the chosen 
realistic $NN$ potential $V^{real}_{NN}$ (see the next section). 

\section{Field-theoretical realization}\label{kap3_kap_ham3}

Now, we will introduce a field-theoretical realization of the hadronic 
interaction. In view of various approaches in the 
literature, it will be useful to distinguish two types of realizations 
which differ in the treatment of the interaction in 
${\cal H}^{[2]}_{\bar N }$ only, i.e., in $V^{con}_{[2]\bar N\bar N}$ 
of (\ref{kap3_z3}) and which we will coin ``static'' and ``retarded'' 
approaches. In order to illustrate the essential differences, we will set 
$V^0_{XX}$ equal to zero for the moment being for simplicity. Then the pure 
nucleonic component $V^{con}_{[2]\bar N\bar N}$ consists of two contributions:
(i) a given hermitean, energy independent potential generated from 
$V^{0\,[2]}_{\bar N\bar N}$, and (ii) a retarded one-meson exchange potential
\begin{equation}\label{kap3_nn_1ret}
V^{ret}_{\bar N \bar N}(z)
= \left[ V_{\bar NX} G_0(z) V_{X\bar N} \right]_{con}\,\,.
\end{equation} 
In the following, we will use the notion {\it static approach} for 
the case that any explicit meson-nucleon vertex  $V^0_{\bar N X}$ vanishes. 
Thus in the static case the retarded 
one-meson exchange $V^{ret}_{\bar N \bar N}$ vanishes identically and 
the $NN$ interaction $V^{con}_{\bar N\bar N}$ is generated by 
$V^{[2]}_{NN}$ alone. One should note, however, that even in the static 
approach retardation is still contained in the interaction 
$V^{con}_{\Delta\Delta}$. 
Furthermore, there is no distinction between a bare 
and a physical nucleon, and the operators ${\widehat Z}_{[2]}$ and 
${\widehat R}$ are both equal to the identity. Consequently, one can 
leave out the ``bar'' in the notation. Moreover, the upper index ``0'' in 
 the interaction operator $V^0$ can be dropped. 
 In order to have a realistic 
description, one then has to {\it identify} $V^{[2]}_{NN} \equiv
 V^{0\,[2]}_{NN}$
with a realistic $NN$ potential model $V^{real}_{N N}$, i.e.,  
\begin{equation}\label{kap3_kraftu}
V^{con}_{[2]N N}(z) = V^{[2]}_{N N}
= V^{real}_{N N} \,.
\end{equation}
Furthermore, one has to keep in mind that due 
to the coupling to the $N \Delta$ and $\pi NN$ states, the realistic potential 
has to be renormalized in order to avoid double counting of parts of the 
interaction as will be discussed below in Sect.~\ref{kap_double_counting}. 
Such an approach has been used in \cite{Sau86,PoS87} and also in the coupled 
channel calculation of \cite{WiA93,WiA96}. The obvious advantage of 
this framework is its simplicity. For excitation energies up to about 500 MeV, 
only pions could be created via the $\pi N\Delta$ vertex. 
 
In the retarded approach, on the other hand, one chooses 
$V^0_{\bar N X} \neq 0$ and $V_{\bar N\bar N}^{0\,[2]}  \equiv 0$. 
Neglecting $V^0_{XX}$ and $V^0_{\Delta \bar N}$ for a moment, the 
$NN$ interaction is generated completely by the retarded one-meson exchange 
part  $V^{ret}_{\bar N \bar N}(z)$. In order to 
obtain a realistic description, one then has to consider besides 
the pion also heavier mesons explicitly. Otherwise, the $NN$ interaction 
would be generated by a retarded one-pion exchange potential 
only, which would result in a rather crude description of experimental data. 
In the present work, we use the potential of Elster 
{\it et al.}~\cite{Els86,ElF88}. Also in this case a renormalization of the 
$NN$ potential will be needed if one includes explicitly the $N \Delta$ and 
$\pi NN$ channels. 
 
It is obvious that besides these two extremes various alternatives are
possible. For example, Bulla and Sauer \cite{BuS92} used in their extension
of the original model \cite{Sau86,PoS87} the choice
\begin{eqnarray}
V^0_{\bar N \pi} \neq 0\,, \quad
V^0_{\bar N x} &=& 0\,, \,\, x \in \{
\rho,\, \omega,\, \sigma,\,  \delta,\, \eta\}\, , \quad 
V^{[2]}_{\bar N  \bar N } = 
V^{real}_{\bar N  \bar N } -
V^{ret}_{\bar N \bar N }(z=2 M_N)\,, \label{kap3_sau3}
\end{eqnarray}
and for $V^{real}_{\bar N  \bar N }$ the Paris 
potential~\cite{LaL80}, which has been renormalized with respect to 
$\pi$ exchange in order to avoid double counting as will be discussed 
in detail below in Sect.~\ref{kap_double_counting}. 

\subsection{The component $V^0_{{\bar N}X}$}
\label{kap3_vqn}

The  components $V^0_{{\bar N}X}$
and $V^0_{X{\bar N}} = \left({V^0_{{\bar N}X}}\right)^{\dagger}$
are explicitly present in {\it retarded} calculations only. For the 
meson-nucleon vertices, we have taken the usual couplings for 
pseudoscalar, scalar, and vector mesons whose explicit forms we have 
taken from \cite{MaH87}. At each vertex we have furthermore introduced a 
phenomenological hadronic form factor $F_{x}(\vec{q}^{\,2})$ parametrized 
in the conventional monopole ($n_x=1$) or dipole ($n_x =2$) form 
\begin{equation}\label{kap3_form}
F_{x}(\vec{q}^{\,2}) 
= \left( \frac{\Lambda_{x}^2 - m_{x}^2}{
\Lambda_{x}^2 + \vec{q}^{\,2}}    \right)^{n_x} \, , \quad
x \in \{ \pi, \eta, \sigma, \delta, \omega, \rho \}\, ,
\end{equation}
where the cutoff parameters $\Lambda_x$ are treated as free parameters 
to be fixed by fitting the $NN$ scattering data below $\pi$ threshold
and the deuteron properties. 

\subsection{The coupling $V^0_{{\Delta} X}$ and the dressing of the 
$\Delta$ isobar}\label{kap3_ndeltavertex}

\begin{figure}[hp]
\centerline{\psfig{figure=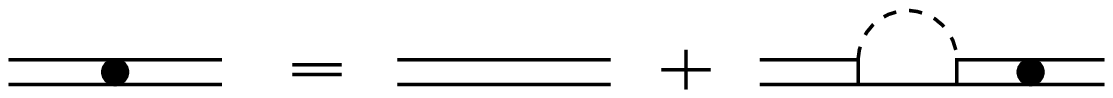,width=10cm,angle=0}}
\vspace{0.5cm}
 \caption{Diagrammatic representation of the dressed $\Delta$ propagator
  $g_{\Delta}$ (see Eq.\  (\ref{kap3_pin5})). In order to distinguish
 $g_{\Delta}$ from the free propagator, it is denoted by a ``$\bullet$''.}
 \label{kap3_abb_gdeltadef}
\end{figure}

In view of the strong coupling of the $\Delta$ isobar to the $\pi N$ system, 
we restrict ourselves to the coupling of the $\Delta$ to the $\pi N$ channel, 
i.e., $V^0_{\Delta {\pi}}\neq 0$ only, for which we take the usual 
nonrelativistic form for the one-body vertices of (\ref{kap3_vdn2})
\begin{eqnarray}  
{v}^{\,0}_{\Delta {\pi}}  (\vec{p}^{\,\prime}\lambda'; 
\vec{p} \, \lambda   \vec{q} \,\mu) &=&
i\,  \frac{f^0_{\Delta N\pi}}{m_{\pi}}\, 
 \chi_{\Delta,\,\lambda'}^\dagger\,{\vec \sigma}_{\Delta {\bar N}} \,\cdot
 \vec{q} \,\chi_{\bar N,\,\lambda}\,  F_{\Delta N\pi}(\vec{q}^{\,2})\,
 \tau_{\Delta \bar N,\,\mu} \,, 
\label{kap3_vertexpind_nr}
\end{eqnarray}
where $\chi_{\Delta,\,\lambda'}$ and $\chi_{\bar N,\,\lambda}$ denote the 
nonrelativistic $\Delta$ and nucleon spinors, respectively, and 
the spin and isospin transition operators 
$\vec \sigma_{\Delta \bar N}$ and $\vec \tau_{\Delta\bar N}$
are fixed by the reduced matrix elements 
$\bra{\frac{3}{2}}|\sigma_{\Delta {\bar N}}^{[1]}|\ket{\frac{1}{2}} 
= \bra{\frac{3}{2}}|\tau_{\Delta\bar N}^{[1]}|\ket{\frac{1}{2}} = 2$.
Again a phenomenological form factor
\begin{equation}\label{kap3_form_pind}
F_{\Delta N\pi}(\vec{q}^{\,2}) 
= \left( \frac{\Lambda_{\Delta N\pi}^2 - m_{\pi}^2}{
\Lambda_{\Delta N\pi}^2 + \vec{q}^{\,2}}   
\right)^{n_{\Delta\pi}}
\end{equation}
has been introduced. Similar to \cite{WiA93,PoS87}, the free parameters 
$f^0_{\Delta N\pi }$ and  $\Lambda_{\Delta N\pi}$ are 
fixed by fitting 
$\pi N$ scattering in the $P_{33}$ channel. One finds for the dressed 
$\Delta$ propagator $g_{\Delta}$, depicted in Fig.~\ref{kap3_abb_gdeltadef}, 
\begin{equation}
 g_{\Delta}(z) = 
 (z-M^0_{\Delta}-\Sigma_{\Delta}(z))^{-1}
 \label{kap3_pin5}
\end{equation}
with the $\Delta$ self energy  
\begin{equation}
 \Sigma_{\Delta}(z) = v^0_{\Delta {\pi}} g_0(z) 
 v^0_{\pi  \Delta}\,. 
 \label{equ:pin6}
\end{equation}
Because of our choice $v^0_{XX}=0$ and the one-meson-approximation,
background mechanisms like, e.g., the Chew-Low term 
have to be neglected as in \cite{WiA93,PoS87}. In detail, one obtains 
for the $\Delta$ mass and width, respectively,
\begin{eqnarray}
 M_{\Delta}(W) &=& M^0_{\Delta} + \frac{4\pi}{3} 
 \PP \int_0^{\infty} 
 \frac{dq'\, q^{\prime\,4}}{(2\pi)^32\omega_{\pi}(q')} 
 \left( \frac{f^0_{\Delta N\pi}}{m_{\pi}} \right)^2
 \frac{F_{\Delta N\pi }^2(q'^2)}{W-
 \omega_{\pi}(q')-e^{nr}_N(q')}\,,
 \label{kap3_pin15}\\
 \Gamma_{\Delta}(W) &=& \left\{ \begin{array}{ll}
 \frac{1}{6\pi} \frac{q^3M_N}{\omega_{\pi}(q)+M_N} 
 \left( \frac{f^0_{\Delta N\pi}}{m_{\pi}} \right)^2
 F_{\Delta N\pi}^2(q^2) \quad & \mbox{for} \quad 
 W > m_{\pi}+M_N \,, \\
  0 & \mbox{for} \quad W \leq m_{\pi}+M_N\,,  \end{array} \right.
 \label{kap3_pin14}
\end{eqnarray}
with the invariant energy $W=\omega_{\pi}(q)+e^{nr}_N(q)$. 
For the free parameters $f^0_{\Delta N\pi}$, $\Lambda_{\Delta N\pi}$ and 
$M^0_{\Delta}$, our fit to the solution SM95 of~\cite{Arn98} yields
\begin{equation}
 \frac{\left(f^0_{\Delta N\pi}\right)^2}{4\pi}=0.9452\,, \quad
 \Lambda_{\Delta N\pi}=482.11\,\mbox{MeV}\,,
 \quad M^0_{\Delta}=1281.7\,\mbox{MeV}\,,
 \quad {n_{\Delta\pi}} =2\,,
 \label{kap3_form2}
\end{equation}
whereas in \cite{WiA93,PoS87}
\begin{equation}
 \frac{\left(f^0_{\Delta N\pi }\right)^2}{4\pi}=1.393\,, \quad
 \Lambda_{\Delta N\pi}=287.9\,\mbox{MeV}\,,
 \quad M^0_{\Delta}=1315\,\mbox{MeV}\,\,, \quad 
 {n_{\Delta\pi}} =1\,\, 
 \label{kap3_form_paul}
\end{equation}
has been used. We would like to emphasize that the values in 
(\ref{kap3_form2}) have been obtained by fitting {\it simultaneously} 
$\pi N$ scattering in the $P_{33}$ channel {\it and} the 
$M_{1+}^{(3/2)}$ multipole of photopionproduction. The reason for 
this procedure will become apparent in a forthcoming paper on e.m.\ 
reactions on the deuteron, because it turned out that a reasonable 
description of the most important $M_{1+}^{(3/2)}$ multipole in 
the $\Delta$ region is not possible in our approach if only $\pi N$ 
scattering is considered for the fit of $f^0_{\Delta N\pi}$, 
$\Lambda_{\Delta N\pi}$ and $M^0_{\Delta}$. 

\subsection{The interactions 
${\bar N}{\bar N}\leftrightarrow {\bar N}\Delta$ 
and ${\bar N}\Delta  \leftrightarrow {\bar N} 
\Delta$}\label{kap3_vnd_vdd}

\begin{figure}[hp]
\centerline{\psfig{figure=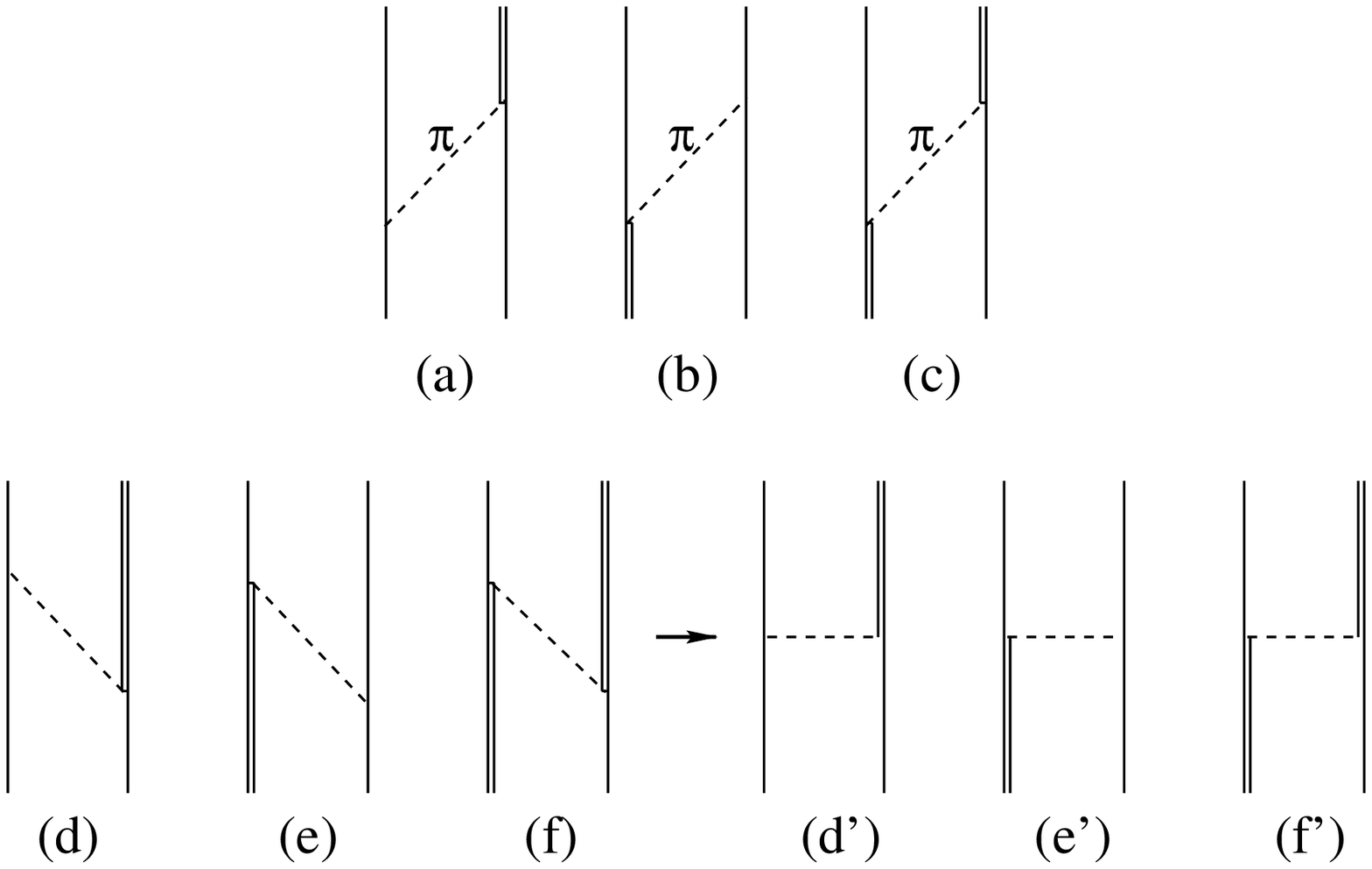,width=11cm,angle=0}}
\vspace{0.5cm}
\caption{One-boson exchange diagrams for the various interactions involving
a ${\bar N} \Delta$ configuration. Diagrams (a) through (c) represent 
retarded contributions (here $\pi$ only). Diagrams (d) through (f) are 
treated in the energy independent approximation as indicated 
by the diagrams (d') through (f') because of the 
absence of $\pi N\Delta$ and $\pi \Delta\Delta$ configurations. }
\label{abb_delta}
\end{figure}

The  time ordered  diagrams of the interactions
${\bar N} \Delta \leftrightarrow {\bar N}{\bar N}$ and 
${\bar N}\Delta \leftrightarrow {\bar N} \Delta$ 
are depicted in Fig.~\ref{abb_delta}. In view of the 
truncation of the model Hilbert space, not allowing explicit 
meson-$\bar N \Delta$ or meson-$\Delta\Delta$ configurations, the 
contributions to the three diagrams (d) through (f) have to be described 
by the ab initio potentials $V^{0\,[2]}_{\Delta \bar N }$, 
$V^{0\,[2]}_{\bar N \Delta}$ and 
$V^{0\,[2]}_{\Delta \Delta}$ in the energy-independent limit, 
represented by the diagrams (d') through (f') in Fig.~\ref{abb_delta}. 
In the static approach, the $\Delta$-$N$ mass difference is neglected in 
the meson propagator of the diagrams (d') through (f') 
while it is retained in the retarded approach.
As in the work of \cite{BuS92,PoS87}, we consider $\pi$ and $\rho$ exchange
only. With respect to the other three time-ordered diagrams (a) through (c) 
of Fig.~\ref{abb_delta}, for which we consider only $\pi$ exchange, one has 
to distinguish retarded and static approaches. Whereas retardation is kept 
for diagram (c) in both cases, diagrams (a) and (b) are only considered 
in the retarded framework. Otherwise they are  contained in 
$V^{0\,[2]}_{\Delta \bar N }$ and $V^{0\,[2]}_{\bar N \Delta}$.

The corresponding matrix elements for retarded $\pi$ exchange have the 
following structure 
\begin{eqnarray}
\bra{{\bar N}{\bar N};\,\vec{p}^{\, \prime}}
[ V^0_{{\bar N}X}\,G_0(z)\,V^0_{X \Delta} ]_{con} 
\ket{{\bar N} \Delta;\,\vec{p}\,} &=&
F_{\pi}(\vec{q}^{\,2})\, F_{\Delta N\pi}(\vec{q}^{\,2})\,
\frac{f^0_{\pi}f^0_{\Delta  N\pi}}{m_{\pi}^2}
\frac{\vec{\tau}_{{\bar N}{\bar N}}(1) \cdot 
\vec{\tau}_{{\bar N}\Delta}(2)}
{(2\pi)^3 2 \omega_{\pi}(q)}\,
\nonumber\\
&&\times \,  
\frac{(\vec{\sigma}_{{\bar N}{\bar N}}(1) \cdot \vec{q}\,)\,
(\vec{\sigma}_{{\bar N}\Delta}(2) \cdot \vec{q}\,)}
{z - e^{nr}_N(p) - e^{nr}_N(p') - 
\omega_{\pi}(q)} +(1 \leftrightarrow 2)\,, \label{kap3_nd_e}\\ 
\bra{{\bar N}\Delta;\,\vec{p}^{\, \prime}}
[V^0_{\Delta X}\,G_0(z)\,V^0_{X\Delta}]_{con} 
\ket{{\bar N}\Delta;\,\vec{p}\,} &=&
 F^2_{\Delta N\pi}(\vec{q}\,^2)\,
 \frac{\left(f^0_{\Delta N\pi}\right)^2}{m_{\pi}^2}
 \frac{\vec{\tau}_{\Delta {\bar N}}(1) 
\cdot \vec{\tau}_{{\bar N} \Delta}(2)}
{(2\pi)^3 2 \omega_{\pi}(q)}\,
\nonumber\\
&&\times   
\frac{(\vec{\sigma}_{\Delta {\bar N}}(1) 
\cdot \vec{q}\,) 
\, (\vec{\sigma}_{{\bar N} \Delta}(2) 
\cdot \vec{q}\,)}
{z - e^{nr}_N(p) - e^{nr}_N(p') - 
\omega_{\pi}(q)} 
+\,(1 \leftrightarrow 2)\,,  \label{kap3_dd_e} 
\end{eqnarray}
where $\vec{q}=\vec{p}^{\,\prime}-\vec{p}$. For the evaluation of these 
expressions, the nonrelativistic reduction of the vertices  
has been used. As in (\ref{kap3_pin15}), 
we take the nonrelativistic nucleon energies $e^{nr}_N(p)$ in the 
$\pi {\bar N}{\bar N}$ propagators for simplicity. 

The remaining one-pion exchange diagrams (d') through (f') in 
Fig.~\ref{abb_delta} and the energy independent limit of {\it all} $\rho$ 
exchange diagrams corresponding to (a) through (f) in Fig.~\ref{abb_delta} are 
included in  $V^{0\,[2]}_{{\bar N} \Delta}$, $V^{0\,[2]}_{\Delta{\bar N}}$ and 
$V^{0\,[2]}_{\Delta \Delta}$, respectively, which we decompose
into $\pi$  and $\rho$ exchange parts according to
 \begin{eqnarray}
 V^{0\,[2]}_{{\bar N} \Delta}(z) &=&
 V^{0\,(\pi)}_{{\bar N} \Delta}(z) + 
 V^{0\,(\rho)}_{{\bar N} \Delta}(z) \, ,\label{kap3_vvoll1}\\
 V^{0\,[2]}_{\Delta \Delta}(z) &=&
  V^{0\,(\pi)}_{\Delta \Delta}(z) + 
 V^{0\,(\rho)}_{\Delta \Delta}(z) \,. \label{kap3_vvoll3}
\end{eqnarray}
Explicitly, we take for them in the retarded approach the 
following expressions 
\begin{eqnarray}
 \bra{{\bar N}{\bar N};\,\vec{p}^{\, \prime}}
 V^{0\,(\pi)}_{{\bar N} \Delta}
 \ket{{\bar N} \Delta;\,\vec{p}\,} &=&
 {F}_{\pi}(\vec{q}\,^2)\, 
 {\bar F}_{\Delta N\pi}(\vec{q}\,^2)\,
 \frac{f^0_{\pi}{\bar f}^0_{\Delta N\pi}}{m_{\pi}^2}
 \frac{\vec{\tau}_{{\bar N}{\bar N}}(1) \cdot 
 \vec{\tau}_{{\bar N}\Delta}(2)}
 {(2\pi)^3 2\omega_{\pi}(q)}\,
 \frac{(\vec{\sigma}_{{\bar N}{\bar N}}(1) \cdot \vec{q}\,)\, 
 (\vec{\sigma}_{{\bar N}\Delta}(2) \cdot \vec{q}\,)}
 {M_N - M^0_{\Delta} - \omega_{\pi}(q)}
 +\,(1 \leftrightarrow 2)\,,  \label{kap3_nd_e2}\\
 \bra{{\bar N}\Delta;\,\vec{p}^{\, \prime}}
 V^{0\,(\pi)}_{\Delta \Delta}
 \ket{{\bar N}\Delta;\,\vec{p}\,} &=&
 {\bar F}^2_{\Delta N\pi}(\vec{q}\,^2)\,
 \frac{\left({\bar f}^0_{ \Delta N\pi}\right)^2}{m_{\pi}^2}\,
 \frac{\vec{\tau}_{\Delta {\bar N}}(1) \cdot 
 \vec{\tau}_{{\bar N} \Delta}(2)}
 {(2\pi)^3 2 \omega_{\pi}(q)}\,
 \frac{(\vec{\sigma}_{\Delta {\bar N}}(1) \cdot \vec{q}\,)\, 
 (\vec{\sigma}_{{\bar N} \Delta}(2) \cdot \vec{q} \,)}
 {2 M_N - 2 M^0_{\Delta} - \omega_{\pi}(q)}
 +\,(1 \leftrightarrow 2)\,.  \label{kap3_dd_e2} 
\end{eqnarray}
The corresponding expressions for the static approach for 
$V^{0\,(\pi)}_{\Delta \Delta}$ are obtained by 
neglecting here the $N$-$\Delta$ mass difference.
The form factor ${\bar F}_{\Delta N\pi}$, parametrized as in 
(\ref{kap3_form_pind}) with parameters $\bar \Lambda_{\Delta N\pi}$ and 
$\bar n_{\Delta \pi}$, and the coupling constant ${\bar f}^0_{ \Delta N\pi}$ 
are {\it not} identical with $F_{\Delta N\pi}$ and $f^0_{\Delta N\pi}$, 
respectively, which were fixed
by fitting $\pi N$ scattering data, where $\pi \Delta$ configurations 
are not considered. Consequently, ${\bar F}_{\Delta N\pi}$ 
and ${\bar f}^0_{ \Delta N\pi}$ can be treated in principle as free 
parameters to be fixed by $NN$ scattering (see the following section). 
 
With respect to $\rho$ exchange, we consider besides the usual 
$\rho {\bar N} \Delta$ interaction density \cite{MaH87} 
\begin{eqnarray}  
-i\frac{{\bar f}^0_{\Delta N\rho}}{m_{\rho}}
{\bar \psi}^{\nu}_{\Delta} \gamma^5 \gamma^{\mu}
{\vec \tau}_{\Delta{\bar N}} \cdot \psi
\left( \partial_{\mu} {\vec \phi}_{\nu} - \partial_{\nu}
 {\vec \phi}_{\mu} \right) + h.c.\,,
  \label{kap3_lagrange1_rnd}
\end{eqnarray}
where $\psi^{\nu}_{\Delta}$, $\psi$, and $\vec \phi_{\nu}$ denote the 
$\Delta$ Rarita-Schwinger spinor, the nucleon Dirac spinor, and the 
$\rho$ meson field, respectively, an additional alternative
 \begin{eqnarray} 
- \frac{{\bar g}^{\,0}_{\Delta N\rho}}{4 M_N^2}
{\bar \psi}^{\mu}_{\Delta} \gamma^5 
{\vec \tau}_{\Delta {\bar N}}
\cdot \left(\partial^{\nu} \psi \right)
\left(\partial_{\mu} {\vec \phi}_{\nu} - \partial_{\nu}
{\vec \phi}_{\mu} \right) + h.\ c.\ \,\,, 
\label{kap3_lagrange2_rnd}
\end{eqnarray}
which, to our knowledge, has not been discussed in the literature.  
A nonrelativistic reduction yields in the c.m.\ frame
\begin{eqnarray}
 \bra{{\bar N}{\bar N};\,\vec{p}^{\, \prime}}
 V^{0\,(\rho)}_{{\bar N} \Delta}
 \ket{{\bar N} \Delta;\,\vec{p}\,} &=&
 \frac{\vec{\tau}_{{\bar N}{\bar N}}(1) \cdot 
 \vec{\tau}_{{\bar N}\Delta}(2) }
 {(2\pi)^3 2 \omega_{\rho}(q)}\,
 \left(\frac{1}{M_N - M^0_{\Delta}
 -\omega_{\rho}(q)}-\frac{1}{\omega_{\rho}(q)}\right)\,
 \nonumber\\
&&\times
 {F}_{\rho }(\vec{q}\,^2)\,
 {\bar F}_{\Delta N\rho}(\vec{q}\,^2)\,
 \frac{{\bar f}^0_{\Delta N\rho}}{m_{\rho}}\,
 \left\{ - \frac{g^0_{\rho}}{2 M_N}\,
 4 i\, \vec{\sigma}_{{\bar N} \Delta}(2)
 \cdot \left( \vec{q} \times \vec{P} \right) 
 \right. \nonumber\\
&&+
 \frac{g^0_{\rho} + f^0_{\rho}}{2 M_N}\,
 (\vec{\sigma}_{{\bar N}{\bar N}}(1) \times \vec{q}\,)\cdot
 (\vec{\sigma}_{{\bar N}\Delta}(2) \times \vec{q}\,) 
 \nonumber\\
& & + \left. 
\frac{g^0_{\rho}}{2M_N} \,\alpha_{\Delta N\rho}\,
(\vec{\sigma}_{{\bar N}{\bar N}}(2)
 \times \vec{q} \,)\cdot
(\vec{\sigma}_{{\bar N} \Delta}(2) \times \vec{q} \,)
 \right\} +\,(1 \leftrightarrow 2)\,,  \label{kap3_rho_nd_e}\\
\bra{\Delta {\bar N};\,\vec{p}^{\, \prime}}
 V^{0\,(\rho)}_{\Delta \Delta}
 \ket{\Delta {\bar N};\,\vec{p}\,} &=&
 \frac{\left({\bar f}^0_{\Delta N\rho}\right)^2}{m_{\rho}^2}
 \, \frac{ \vec{\tau}_{\Delta {\bar N}}(1) \cdot 
 \vec{\tau}_{{\bar N}\Delta}(2)}
 {(2\pi)^3 2 \omega_{\rho}(q)}
 \left( \frac{1}{2 M_N - 2 M^0_{\Delta} - \omega_{\rho}(q)} 
 -\frac{1}{\omega_{\rho}(q)}\right)
\nonumber\\
&&\times  
 {\bar F}^2_{\Delta N\rho}(\vec{q}\,^2)\, 
 (\vec{\sigma}_{\Delta {\bar N}}(1) \times \vec{q}\,)\cdot
 (\vec{\sigma}_{{\bar N} \Delta}(2) \times \vec{q}\,) 
+(1 \leftrightarrow 2)\,   \label{kap3_rho_dd_e} 
\end{eqnarray}
with
\begin{equation}\label{kap3_defpq}
\vec{q} = \vec{p}^{\,\prime} - \vec{p}\, ,
 \qquad 
\vec{P} = \frac{1}{2} \left(\vec{p}^{\,\prime} + \vec{p}\,\right)\,,
\end{equation}
and
\begin{equation}
\alpha_{\Delta N\rho}=1-\frac{m_{\rho}}{4 M_N}
\frac{{\bar g}^{\,0}_{\Delta N\rho}}{{\bar f}^0_{\Delta N\rho}}
\,.\label{alpha_rho}
\end{equation}
Again the form factor ${\bar F}_{\Delta N\rho}(\vec{q}\,^2)$ and
coupling constant ${\bar f}^0_{\Delta N\rho}$ are fitted 
to $NN$ scattering, whereas $f^0_{\rho}$ are  $g^0_{\rho}$ are fixed by
the values of the Elster-potential. Note that the terms proportional to 
$( \vec{q} \times \vec{P} )$ and $(\vec{\sigma}_{{\bar N}{\bar N}}
(2) \times \vec{q} \,)\cdot(\vec{\sigma}_{{\bar N} \Delta}(2) 
\times \vec{q} \,)$ are usually neglected in the literature, 
for example in \cite{BuS92,PoS87}.

Finally, with respect to the static approach, only the  
${\bar N} \Delta$ potential (\ref{kap3_dd_e}), represented by the diagram (c) 
in Fig.~\ref{abb_delta}, is generated by iteration of the 
$\pi {\bar N} \Delta$ vertex, as has already been mentioned above. 
All other diagrams of Fig.~\ref{abb_delta} have to be described by 
$V^{0\,[2]}_{PP}$ where in addition the $N$-$\Delta$ mass difference 
is neglected. Consequently, we obtain in this limit 
\begin{eqnarray} 
 \bra{N N;\,\vec{p}^{\, \prime}}
 V^{0\,(\pi)}_{N \Delta}
\ket{N \Delta;\,\vec{p}\,} &=&
 - F_{\pi}(\vec{q}\,^2)\, {\bar F}_{\Delta N\pi}(\vec{q}\,^2)
 \frac{f^0_{\pi} {\bar f}^0_{\Delta N\pi}}{m_{\pi}^2}\,
 \frac{\vec{\tau} _{NN}(1) \cdot \vec{\tau} _{N\Delta}(2) }
 {(2\pi)^3  \omega^2_{\pi}(q)} \,
\nonumber\\ & & \times  (\vec{\sigma} _{NN}(1)
 \cdot \vec{q} \,)\cdot (\vec{\sigma} _{N\Delta}(2) \cdot \vec{q} \,)
 +(1 \leftrightarrow 2)\,,   \label{kap3_nd_e2_stat} 
\end{eqnarray}
whereas 
Eqs.~(\ref{kap3_dd_e2}), (\ref{kap3_rho_nd_e}), and (\ref{kap3_rho_dd_e}) 
apply also to the static limit except for the neglect of the $N$-$\Delta$ 
mass difference. Note however, that in this case {\it all} coupling 
constants and cutoffs in $V^0_{N \Delta}$ and $V^0_{\Delta \Delta}$
can be treated in principle as free parameters due to the choice
 $V^0_{{\bar N}X}  = 0$.

\subsection{The interaction $V^0_{XX}$}\label{kap3_vqq_uni}

The simplest choice for the diagonal interaction in the subspace 
${\cal H}^{[2]}_{X}$ would certainly be 
$V^0_{XX}\equiv 0$. However, such a choice would lead to severe 
inconsistencies with respect to pion d.o.f., 
because it would lead to a violation of three-body unitarity. The 
reason for this violation lies in the fact
that for excitation energies up to about 500 MeV, $\pi NN$  and 
$\pi d$ states can exist in this sector as asymptotically free states 
($2\pi$  and $3\pi$ states are not allowed due to the 
one-meson-approximation). 
It is therefore obvious, that the interaction $V^{0\,\pi}_{XX}$, which 
describes the $NN$ interaction in the presence of a spectator pion, 
must be considered at least in the $^3S_1$-$^3D_1$ channel which we will 
henceforth refer to as the $\pi d$ channel. Otherwise, the $\pi d$ state 
would not be present formally and reactions like $\pi d \rightarrow \pi d$ 
could not be studied without creating inconsistencies. In 
\cite{PeG92,WiA95,WiA96}, 
for example, reactions with a $\pi d$ state in the initial and/or final 
state were studied without considering $V^{0\,\pi}_{XX}\neq 0$ leading to
a violation of unitarity above the $\pi d$ threshold. 
On the other hand, a $\pi d$ state could not be generated 
if  $V^{0\,\pi}_{XX} =0$. Therefore, 
$V^{0\,\pi}_{XX}$ has to be nonzero, since we are interested 
in the construction of a unitary model up to the $2\pi$ threshold. All other 
diagonal interactions in ${\cal H}^{[2]}_{X}$ do not affect 
three-body unitarity for energies up to 500 MeV and, therefore, can safely 
be set equal to zero for the sake of simplicity
(see (\ref{kap3_vqq}) and (\ref{kap3_vqq_x})). We consider this choice as 
the minimal requirement in order to satisfy three-body unitarity.

\begin{figure}[hp]
\centerline{\psfig{figure=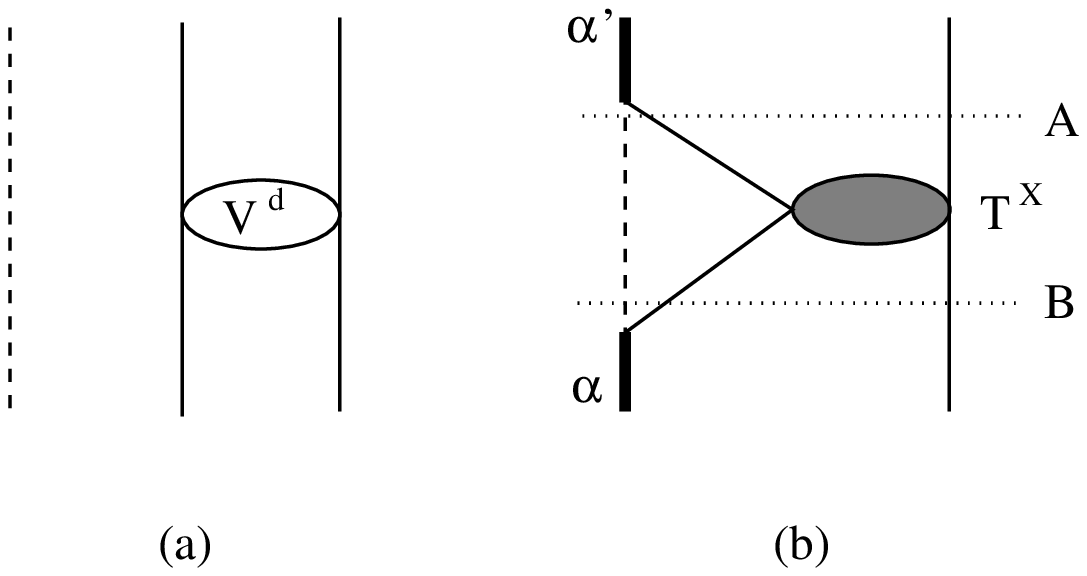,width=9cm,angle=0}}
\vspace{0.5cm}
\caption{Diagrammatic representation of (a) $V^0_{XX}$ describing the 
interaction $V^d_{NN}$ between the two nucleons in the $^3S_1$-$^3D_1$ channel 
with a pion as spectator, and (b) of the term 
$[V_{PX}\,G_0(z)\,T^X(z)\,G_0(z)\,V_{XP}]_{con}$ 
in Eq.~(\ref{kap3_z3}).}
\label{abb_vqq}
\end{figure}

Thus we retain in $V^0_{XX}$ solely the interaction of two nucleons with 
isospin $t=0$ in the presence of a pion as spectator, where we restrict 
ourselves to an interaction, called $V^d$, which acts only in the 
$^3S_1$-$^3D_1$ channel. Then $V^0_{XX}$ can be written in the form 
(see Fig.~\ref{abb_vqq})
\begin{equation}\label{kap3_vqq10}
V^0_{XX} =  \int \frac{d^3q}{(2\pi)^3 2 \omega_{\pi}(q)}
 \ket{\pi(\vec{q}\,)} V^d(M_d + \frac{\vec{q}^{\,2}}{4 M_N}) 
\bra{\pi(\vec{q}\,)}\,,
\end{equation}
where $\ket{\pi(\vec{q}\,)}$ denotes the spectator pion state
with momentum $\vec{q}$. For practical reasons, we use for 
$V^d$ a separable interaction  of rank~1
\begin{equation}
V^d(\alpha)  = \frac{G_0^{-1}(\alpha) \ket{d}\bra{d}
  G_0^{-1}(\alpha)}{
\bra{d}  G_0^{-1}(\alpha) \ket{d}} \,, 
\label{kap3_vqq20}
\end{equation}
which satisfies the Schr\"odinger equation for the nucleonic component
of the deuteron 
\begin{equation}
\left( H_0 + V^d(M_d) - M_d\right)  \ket{d} = 0\,.
\end{equation}
The separable structure of $V^d$ leads to a rather simple expression 
for the relevant amplitude $T^X$ of (\ref{kap3_tX_mod}). One obtains
 \begin{equation}\label{kap3_vqq28}
T^X(z) =  \int \frac{d^3q}{(2\pi)^3 2 \omega_{\pi}(q)}
 \ket{\pi(\vec{q}\,)} T^d(M_d+ \frac{\vec{q}^{\,2}}{4 M_N},
z- \frac{\vec{q}^{\,2}}{4 M_N} - \omega_{\pi}(q))
\bra{\pi(\vec{q}\,)} \,,
\end{equation}
where the amplitude $T^d$ is given  by the analytic expression
\begin{equation}\label{kap3_t^d_sep}
T^d(\alpha,z)   = \frac{G_0^{-1}(\alpha) \ket{d}\bra{d}
  G_0^{-1}(\alpha)}{\bra{d}  G_0^{-1}(\alpha) \ket{d}
- \bra{d}  G_0^{-1}(\alpha)G_0(z)G_0^{-1}(\alpha)    
\ket{d}} \,.
\end{equation}
The shifted arguments in $T^d$ in (\ref{kap3_vqq28}) have their
origin in the fact that the two nucleons, interacting via $V^d$, have a 
total momentum $-\vec{q}$. We use the nonrelativistic energy in order 
to separate the c.m.\ energy exactly. It is therefore natural to use the 
nonrelativistic energy also in the propagator $G_0$ in (\ref{kap3_t^d_sep}). 
Boost contributions to the deuteron wave function are expected to be small 
\cite{GoA92} and thus are neglected. In our numerical evaluation, the 
intermediate $\pi\bar N\bar N$ propagation, entering into the
term $[V_{PX}\, G_0(z)\,T^X(z)\, G_0(z)\,V_{XP}]_{con}$
(cut A and B in Fig.~\ref{abb_vqq}) of Eq.~(\ref{kap3_z3})
is treated in the static  limit for simplicity. Moreover, we use for the
$\pi {\bar N}$ vertex $V^0_{{\bar N}X}$ in 
$[V^0_{PX}\, G_0(z)\, T^X(z)\, G_0(z)\, V^0_{XP}]_{con}$ the 
nonrelativistic version. Both approximations lead to
a slight violation of unitarity which, however, is not critical.

\begin{figure}[hp]
\vspace{0.5cm}
\centerline{\psfig{figure=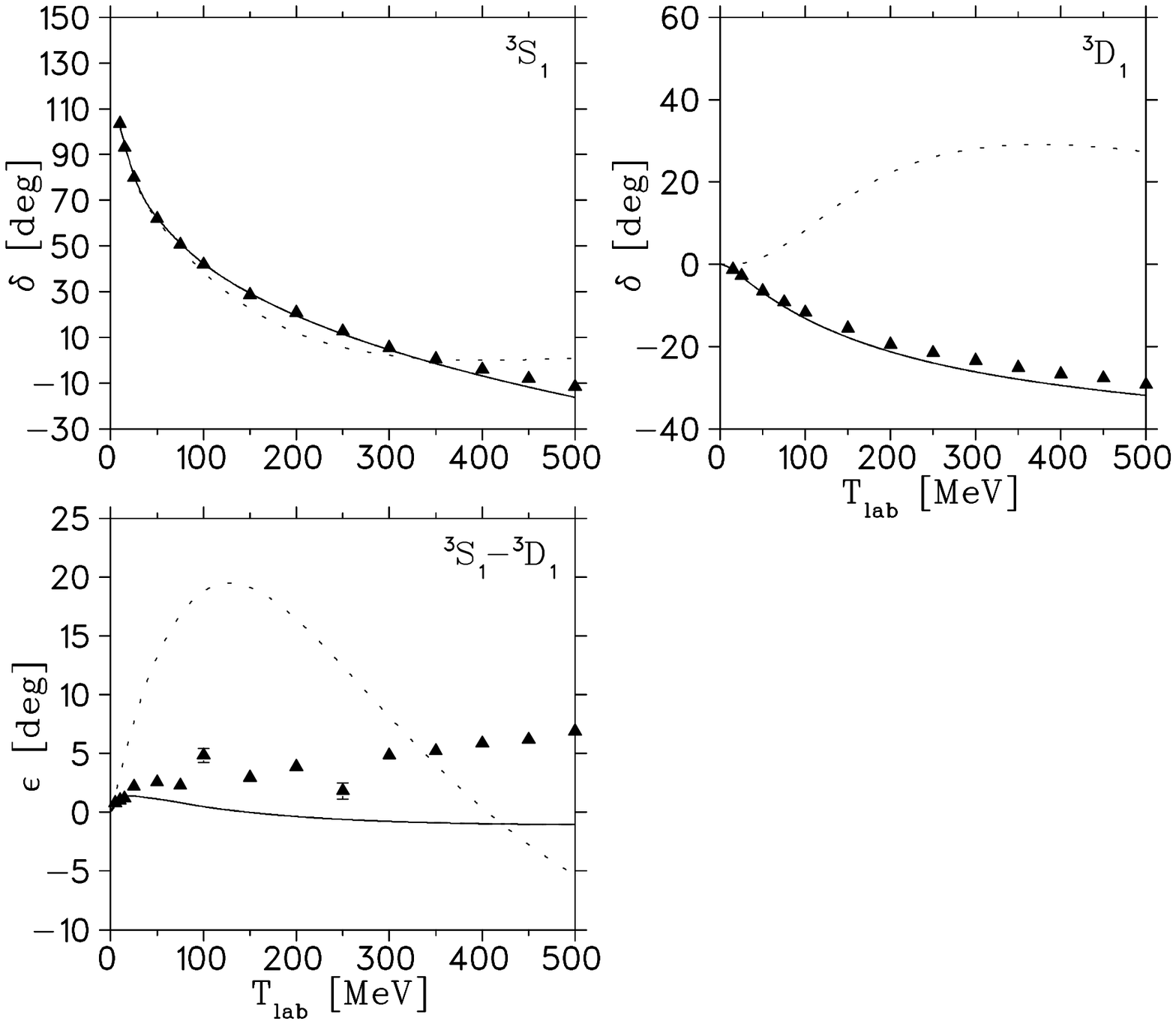,width=10cm,angle=0}}
\vspace{0.5cm}
\caption{Phase shifts and mixing angle for the $^3S_1$-$^3D_1$ channel of 
$NN$ scattering as a function of the lab kinetic energy  $T_{lab}$, compared 
with the experimental data (solution SM97 of Arndt {\it et al.} 
{\protect\cite{Arn98}}). Notation of the curves: dotted: separable ansatz 
({\protect\ref{kap3_vqq20}}) with the deuteron
wave function of the Bonn OBEPR potential;
solid: exact calculation using the full OBEPR potential.}
\label{abb_3s13d1_sep}
\end{figure}

At the end of this subsection, we will discuss briefly the quality of the 
separable interaction in the $^3S_1$-$^3D_1$ channel. For the deuteron pole, 
i.e., for  $W=M_d$, the resulting amplitude  $T^d$ is identical to 
the exact solution obtained with the Bonn-OBEPR  and Elster-potentials, 
respectively. However, for $W \neq M_d$ this equivalence breaks down. 
Concerning the resulting phase shifts, one obtains a strong deviation from 
the exact calculation for the $^3D_1$ phase shift and the mixing parameter 
$\epsilon$ (see Fig.~\ref{abb_3s13d1_sep}). The most important 
$^3S_1$ channel, however, is described reasonably well. These facts indicate 
the limits of the separable ansatz.

\subsection{Renormalization of the realistic $NN$ potential}
\label{kap_double_counting}

\begin{figure}[hp]
\centerline{\psfig{figure=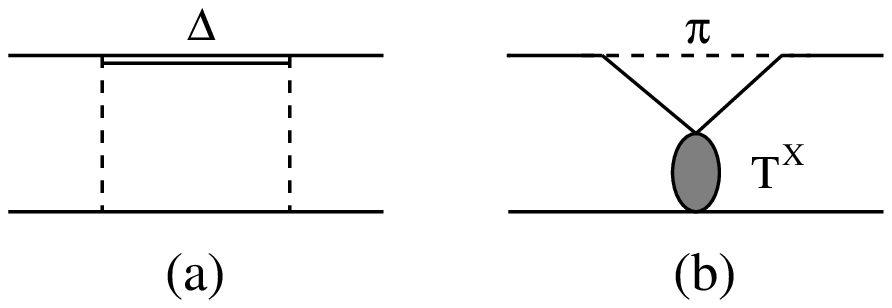,width=8cm,angle=0}}
\vspace{0.5cm}
\caption{Dispersive contributions to the $NN$ interaction:
(a) from intermediate $N\Delta$ states, (b) from intermediate offshell 
$NN$ scattering in the presence of a spectator meson.}
\label{abb_ndbox}
\end{figure}

A problem of double counting in the nucleon-nucleon interaction appears if 
one starts from a potential incorporating effectively certain 
d.o.f., which have been projected out before but which are introduced 
again explicitly later on. For example, let us consider a realistic $NN$ 
potential like Bonn-OBEPR, Bonn-OBEPQ~\cite{Mac89,MaH87}, Paris~\cite{LaL80}, 
Argonne $V_{14}$~\cite{WiS84}, or Nijmegen~\cite{StK93}. These 
potentials act in pure nucleonic space and are fitted to deuteron properties 
and $NN$ scattering data below $\pi$ threshold. 
However, if such potentials are used in a model with explicit 
$\Delta$ d.o.f.\ within a coupled channel approach, the problem 
of double counting becomes evident, because, for example, the dispersive 
box graphs depicted in Fig.~\ref{abb_ndbox} and implicitly present already in 
$V_{NN}$ would be considered explicitly in addition. 

A simple way out of this problem is the box renormalization
of Green and Sainio \cite{GrS82} which consists in a subtraction of 
the box diagrams at a fixed energy $E_0$ from $V_{\bar N \bar N }^{0,\,con}$
in (\ref{kap3_vcon_mod}). We will adopt this method also here and 
use -- similar to previous 
work \cite{WiA93,BuS92,Sau86,PoS87,WiA96} -- the value $E_0 = 2 M_N$. 
With respect to the above discussion we will distinguish

(i) static calculations 
 ($V^0_{\bar N X}=0$):
 \begin{eqnarray}
V^{0\,[2]}_{NN}&=& V^{real}_{NN} - \left.V^{0}_{N\Delta} 
G_0(z) V^{0}_{\Delta N}\right|_{z=E_0}\label{kap3_renvoll1} 
- \left.V^0_{N X} G_0(z) T^X(z) G_0(z) V^0_{X N} \right|_{z=E_0}  \,, 
\end{eqnarray}
and 

(ii) retarded calculations ($V^0_{\bar N X}\neq 0$, 
 $V^{real}_{\bar N \bar N }  =0$):
\begin{eqnarray}
V^{0\,[2]}_{\bar N \bar N }&=& - \left.V^{0\,full}_{\bar N \Delta}(z) 
G_0(z) V^{0\,\,  full}_{\Delta \bar N }(z)\right|_{z=E_0}
-\left.V^0_{\bar N X} G_0(z) T^X(z) G_0(z) 
V^0_{X\bar N } \right|_{z=E_0} \,, 
\label{kap3_renvoll_ret1} 
\end{eqnarray}
where we have defined
\begin{eqnarray}
\label{kap3_ndelta_full} 
V^{0\,\,  full}_{\bar N\Delta}(z)&=&V^{0}_{\bar N\Delta}(z)
+\left[V^0_{\bar N X} G_0(z)V^0_{X\Delta} \right]_{con}\,.
\end{eqnarray}

It should be emphasized that the intermediate $\bar N\Delta$
and $\pi d$ states have isospin $t=1$ so that the box subtraction
has no influence on the $(t=0)$ channels, especially for the deuteron. 
However, such a subtraction would appear if also $\Delta\Delta$ configurations 
would be included explicitly. Furthermore, we would like to remark, that 
this subtraction is not a fundamental ingredient of our approach. It is 
only a relatively simple recipe in order to incorporate a given realistic 
$NN$ potential, which does not contain explicit $\Delta$ degrees of freedom, 
into a $N \Delta$ coupled channel approach, without a complete refit 
of all potential parameters. Obviously, the box subtraction  
will be obsolete in the future by the construction of a realistic 
potential which incorporates from the beginning explicitly nucleon and 
$\Delta$ d.o.f.\ in a coupled channel approach. Finally, we 
would like to remark that the inclusion of (\ref{kap3_renvoll_ret1}) in the 
Lippmann-Schwinger equation (\ref{kap3_tpp3_mod}) leads to an energy 
dependence of the subtracted box graphs which we avoid by the 
substitution in (\ref{kap3_tpp3_mod})
\begin{equation}\label{kap3_ersetzung}
{\widehat R}(z) {V}^{0\,[2]}_{\bar N \bar N }{\widehat R}(z) \rightarrow
\{{\widehat R}(z) {V}^{0\,[2]}_{\bar N \bar N }{\widehat R}(z)\}|_{z=E_0}\,.
\end{equation} 

\subsection{The nucleon-nucleon interaction}\label{kap3_nn_ww}

The properties of the two-nucleon system is governed by the $NN$ interaction. 
In this work, we restrict ourselves to realistic $NN$ potentials known from 
the literature, which have to be renormalized according to the 
discussion in the foregoing subsection. 
The starting points of our considerations are the Lippmann-Schwinger 
equation for the $NN$ scattering matrix (\ref{kap3_tpp3_mod}) and the 
Schr\"odinger equation for the effective, purely nucleonic component of 
the deuteron (\ref{kap3_zwischen3}), which we write in the forms, respectively,
\begin{eqnarray}
T^{con}_{{\bar N}{\bar N}}(z) &=&  
V^{eff}_{{\bar N}{\bar N}}(z) +
V^{eff}_{{\bar N}{\bar N}}(z) 
  G_0(z) T^{con}_{{\bar N}{\bar N}}(z)\,,\label{kap3_tpp3_k}\\
\left( H_0 + V^{eff}_{{\bar N}{\bar N}}(M_d) \right) \ket{d} &=& M_d \ket{d}\,,
\label{kap3_zwischen3_k}
\end{eqnarray}
with the effective $NN$ interaction  
 \begin{equation}\label{kap3_tildev}
V^{eff}_{{\bar N}{\bar N}}(z) =
 {\widehat R}(z) \left( V^{[2]}_{{\bar N}{\bar N}} +
\left[V_{{\bar N}X} G_0(z) V_{X{\bar N}}  \right]_{con} 
\right){\widehat R}(z) \,. 
\end{equation}
Note that in (\ref{kap3_tpp3_k}) the $\Delta$ isobar and the $\pi d$ channel 
has been neglected, because the realistic $NN$ potential models 
do not consider these additional degrees of freedom in general. Exceptions
are, for example, the ``full'' Bonn and the Argonne 
$V_{28}$ potentials \cite{MaH87,WiS84} which include the $\Delta$.
 
In view of the discussion in the foregoing subsection, 
we follow different strategies in static
and retarded calculations. In the static limit the effective interaction 
simplifies to 
\begin{equation}\label{kap3_vtilde1}
V^{eff}_{NN}(z) =V^{[2]}_{NN} \,,
\end{equation}
for which we use the Bonn-OBEPR potential  $V^{OBEPR}$
\cite{MaH87} to be renormalized according to Eq.\ (\ref{kap3_renvoll1}),
whereas in the retarded approach $V^{[2]}_{{\bar N}{\bar N}}$ in 
$V^{eff}_{{\bar N}{\bar N}}(z)$ has to be chosen as 
(cf.~(\ref{kap3_renvoll_ret1})):
\begin{eqnarray}
V^{[2]}_{{\bar N}{\bar N}} &=& ({\widehat Z}^{os}_{[2]})^{-1}
V^{0\,[2]}_{{\bar N}{\bar N}}
({\widehat Z}^{os}_{[2]}) ^{-1}
\label{kap3_v2_ren}\\ 
&=& - \left.\bar V_{{\bar N}\Delta}(z) 
G_0(z) \bar V_{\Delta {\bar N}}(z)\right|_{z=2M_N}
- \left.V_{{\bar N} X}\, G_0(z)\,T^X(z)\, G_0(z) \,
V_{X{\bar N}} \right|_{z=2M_N} \,,
\nonumber 
\end{eqnarray}
with  
\begin{eqnarray}
\bar V_{{\bar N} \Delta}&=&{\widehat Z}_{[2]}^{os} 
V^{0\,full}_{{\bar N} \Delta} ({\widehat Z}_{[2]}^{os})^{-1} 
\label{kap3_voll1}
\end{eqnarray}
(note the analogy to (\ref{kap3_renorm_mat2_5_mod})). 
Concerning the one-boson exchange part in (\ref{kap3_tildev}), we are able 
to use the potential model $V^{Elster}$ of Elster 
{\it et al.}~\cite{Els86,ElF88}, which can be considered as an extension 
of the retarded Bonn-OBEPT potential \cite{MaH87} with respect to the 
inclusion of additional $\pi {\bar N}$ loops. Consequently, we use
in the retarded calculations (note (\ref{kap3_ersetzung}))
\begin{equation}\label{kap3_obept_ren}
V^{eff}_{{\bar N}{\bar N}}(z) = V^{Elster}(z)
 +\left.{\widehat R^{\pi}}(z) V^{[2]}_{{\bar N}{\bar N}}
{\widehat R^{\pi}}(z)\right|_{z=2M_N}\, ,
\end{equation}
where  $V^{[2]}_{{\bar N}{\bar N}}$ is given by 
(\ref{kap3_v2_ren}), and the pionic part ${\widehat R^{\pi}}(z)$ of the 
dressing factor is discussed below. 

Because of its importance for the present work, it is worthwhile to study 
$V^{Elster}$ in some more detail. Concerning the one-boson exchange, 
all six mesons $\pi,\, \eta,\, \sigma, \,\delta,\, \omega$ and $\rho$ 
are taken into account whereas only the pion is considered in the dressing 
operator, therefore denoted by $\widehat R^\pi(z)$, for simplicity. This 
approximation is, however, not crucial with respect to unitarity up to 
the $\eta$ threshold. Consequently, $V^{Elster}(z)$ has the structure
\begin{equation}\label{kap3_elster1}
V^{Elster}(z) = 
{\widehat R^{\pi}}(z) \sum_{x\in \{\pi, \eta, \sigma, \delta, \omega, \rho\}}
\left[V_{{\bar N}x}\, G_0(z)\, V_{x{\bar N}} 
\right]_{con}{\widehat R^{\pi}}(z)\,, 
\end{equation}
where $\widehat R^\pi(z)$ is given in analogy to (\ref{kap3_renorm_mat_r1}) 
with only $\pi N$ loops in ${\widehat Z}^{\pi}_{[2]}(z)$. 
Explicitly, one obtains for the matrix element of the latter 
%using the  ${\pi {\bar N}}$ vertex of (\ref{kap3_vertexpi})
\begin{eqnarray}\label{kap3_z2_explizit}
\bra{{\bar N}{\bar N};\,\vec{p}^{\,\prime}}{\widehat Z}^{\pi}_{[2]}(z)
\ket{{\bar N}{\bar N};\,\vec{p}\,}
&=& \delta \left(\vec{p}\,' - \vec{p}\,\right) {Z}^{\pi}_{[2]}(z,p)
 \label{kap3_z2_m} 
\end{eqnarray}
with
\begin{eqnarray}
\left({Z}^{\pi}_{[2]}(z,p)\right)^2
 &=&1 + 2 \, \frac{\left(g^0_{\pi}\right)^2}{4 \pi} \frac{3}{4 \pi 
e_N(p)} \int_0^{\infty} dk \frac{k^2}{e_N(k)}
\int_{-1}^{1} d \cos \theta\,
 F^2_{\pi}((\vec{p}-\vec{k}\,)^{\,2}) 
\nonumber \\  
& &\times \frac{e_N(p) e_N(k) - M_N^2
- p k  \cos \theta}{ \omega_{\pi}(\vec{p}-\vec{k}\,) \left(
e_N(p) - e_N(k) -  \omega_{\pi}(|\vec{p}-\vec{k}\,|)\right)
\, \left(z - e_N(k) -  \omega_{\pi}(|\vec{p}-\vec{k}\,|)\right)
}\, , \label{kap3_z_e}
\end{eqnarray}
where the  angle between $\vec{p}$ and $\vec{k}$ is denoted  by 
$\theta$.

\begin{figure}[hp]
\centerline{\psfig{figure=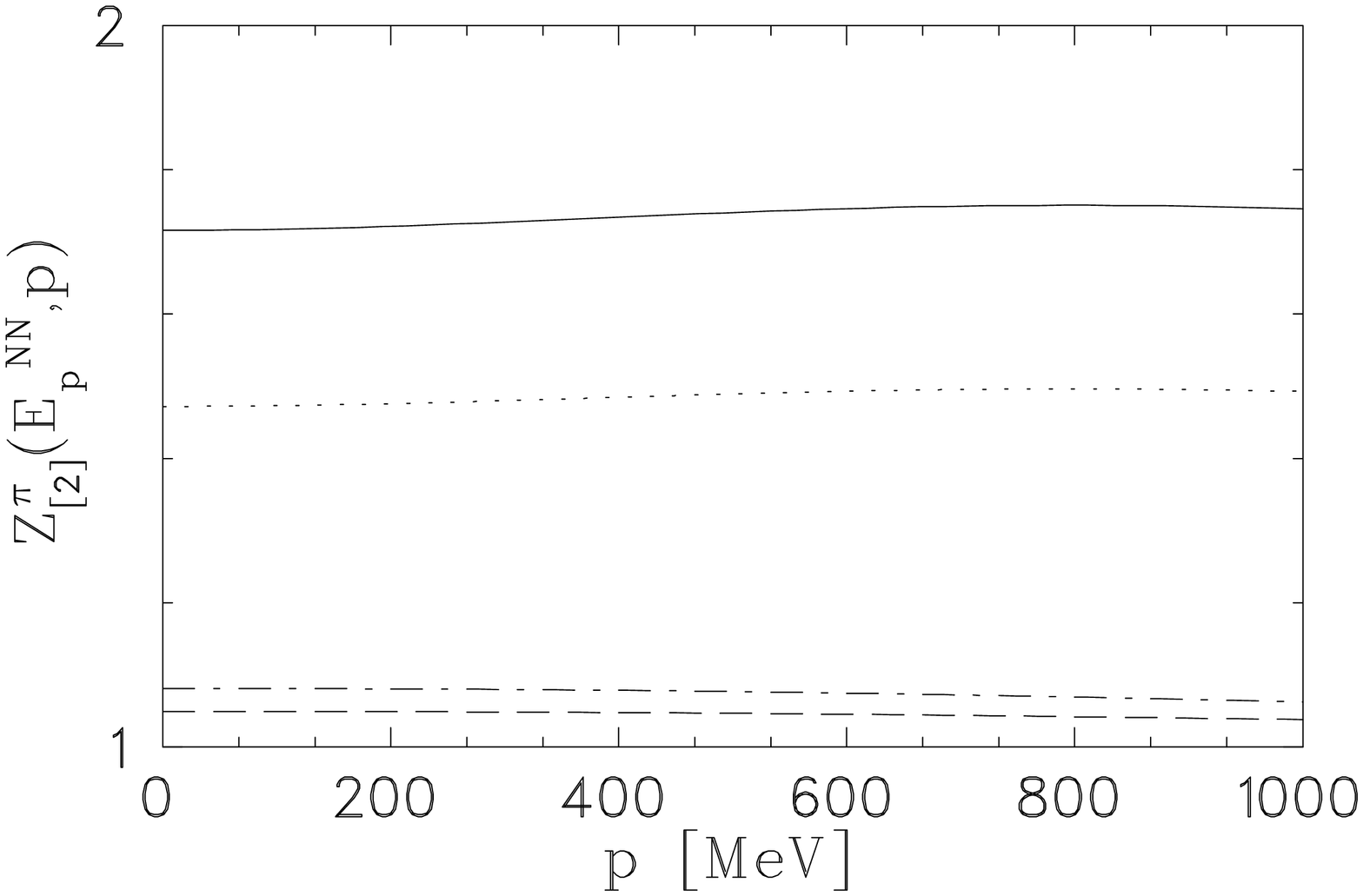,width=9cm,angle=90}}
\vspace{0.5cm}
\caption{ 
The renormalization factor 
$Z^{\pi}_{[2]}(E^{NN}_{p},\,p)$ of Eq.\  
 ({\protect\ref{kap3_z2_explizit}}) as function of the external
 nucleon momentum $p$ for different parameter sets:
 dashed:
 $\frac{{\left(g^0_{\pi}\right)}^2}{4 \pi} = 15$, $\Lambda_\pi =500$ MeV;
 dash-dotted:
 $\frac{{\left(g^0_{\pi}\right)}^2}{4 \pi} = 15$, $\Lambda_\pi =1700$ MeV;
 dotted: $\,\,\frac{\left({g^0_{\pi}}\right)^2}{4 \pi} =
 25$, $\Lambda_\pi =500$  MeV;
 solid: $\,\,\frac{{\left(g^0_{\pi}\right)}^2}{4 \pi} = 25$,
 $\Lambda_\pi =1700$ MeV. A dipole form factor has been used.}
\label{abb_z2}
\end{figure}

In view of the renormalization of $V^0_{{\bar N}X}$ by the operator 
$\widehat Z^{\pi}_{[2]}$, renormalized meson-nucleon coupling constants 
 \begin{eqnarray}
f_x &=& \frac{f^0_x}{{Z}^{\pi}_{[2]}(E_p^{NN},p)} \, \mbox{ and }\,\,
g_x = \frac{g^0_x}{{Z}^{\pi}_{[2]}(E_p^{NN},p)} \label{kap3_kopp_2}
\end{eqnarray}
will appear in (\ref{kap3_elster1}), whose momentum dependence is small and, 
therefore, can be neglected (see Fig.~\ref{abb_z2}). The free parameters in 
(\ref{kap3_elster1}), i.e., the cutoffs and physical coupling constants 
are fixed by fitting $NN$ scattering data below $\pi$ threshold and deuteron
properties. The resulting values are presented in Table~\ref{kap3_tab_cutoff}. 

\section{Results for $NN$ scattering}\label{kap6}

\subsection{Determination of the parameters}\label{kap6_p}

Now we will fix the remaining free parameters in the hadronic interaction 
by considering $NN$ scattering. As discussed in 
Sect.~\ref{kap3_vnd_vdd}, in a retarded approach the parameters of the 
diagrams (a) through (c) of Fig.~\ref{abb_delta} are in principle
fixed  by the parametrization of the meson-nucleon vertices 
(Sect.~\ref{kap3_vqn}) and of the $\pi {\bar N} \Delta$ vertex
$V^0_{{\bar N}\Delta}$ (Sect.~\ref{kap3_ndeltavertex}).
Consequently, only the cutoffs and coupling constants in the diagrams
(d') through (f') can be treated as free parameters. According to 
(\ref{kap3_vvoll1}) through (\ref{alpha_rho}), these diagrams contain
five open parameters, namely the coupling constants 
${\bar f}^0_{\Delta N\pi}$ and ${\bar f}^0_{\Delta N \rho}$, 
the parameter $\alpha_{\Delta N\rho}$ and the form factors
${\bar F}_{\Delta N\pi}(\vec{q}\,^2)$ and 
${\bar F}_{\Delta N \rho}(\vec{q}\,^2)$ which are parametrized as in 
(\ref{kap3_form}).
However, the question arises whether the parametrization of the 
$\pi {\bar N} {\bar \Delta}$ vertex as obtained from fitting $\pi N$ scattering
data should be used for the OBE mechanisms, too. For example, one of the 
differences between $NN$  and $\pi N$ scattering is the fact that in 
$NN$ scattering above $\pi$ threshold a pion {\it can} be onshell, whereas 
in $\pi N$ scattering it {\it must} be onshell. This leads to dramatic 
differences in the cutoff ${\bar \Lambda}_{\Delta N \pi}$ in 
(\ref{kap3_form_pind}). For example, the value of 1200 MeV (with 
${\bar n}_{\Delta\pi} =1$) for 
${\bar \Lambda}_{\Delta N \pi}$ in the full Bonn potential \cite{MaH87} 
is much stronger than the value of typically 300 MeV 
(${\bar n}_{\Delta\pi} =1$) or 500 MeV  (${\bar n}_{\Delta\pi} =2$) 
obtained from $\pi N$ scattering. Similar differences do occur also for 
the parametrization of the $\pi N$ vertex in $NN$  versus $\pi N$ scattering. 

Holzwarth and Machleidt \cite{HoM97} have shown in a detailed analysis that
these problems can be traced back to the use of the  monopole or dipole
parametrization of the hadronic form factors, 
which ties together the low and high 
momentum behaviour and thus is not able to describe both $NN$  and
$\pi N$ data simultaneously with the {\it same} cutoff. In $NN$ scattering,
for example, pion momenta in the neighbourhood of $q=0$ are the relevant 
ones. Thus for small $q$, the form factor should be close 
to unity in order to achieve, for example, a quantitative description of the 
deuteron quadrupole moment, which is dominated by long-range mechanisms.   
Therefore, a large cutoff mass is needed. 
On the other hand, for large cutoff masses the form factor does not 
decrease fast enough with increasing pion momenta as is 
necessary in $\pi N$ scattering. However, a form factor which is derived 
from the Skyrme model is able to describe both {\it simultaneously}, $NN$  as well as 
$\pi N$ scattering \cite{HoM97}, because it combines 
the features of a hard monopole or dipole form factor at $q=0$ with the 
quality of a soft form factor for larger pion momenta. 

For these reasons -- although we are aware of the fact that this procedure 
leads to additional inconsistencies --,
we do {\it not} use the values (\ref{kap3_form2}) in the OBE mechanism 
of our coupled ${\bar N}{\bar N}$-${\bar N} \Delta$ system. 
Thus, we replace the coupling constant $f^0_{\Delta N\pi}$ and the form 
factor $F_{\Delta N\pi}(\vec{q}^{\,2})$ in the retarded diagrams (a) and (b) 
of Fig.~\ref{abb_delta} (see (\ref{kap3_nd_e})) by the corresponding 
quantities ${\bar f}^0_{\Delta N\pi}(\vec{q}\,^2)$ and 
${\bar F}_{\Delta N\pi}(\vec{q}\,^2)$ of diagram (d') and (e'), respectively. 
For the ${\bar N} \Delta \leftrightarrow {\bar N} \Delta$ transition
(diagram (c) and (f') in Fig.~\ref{abb_delta}) we use the values 
(\ref{kap3_form2}) or (\ref{kap3_form_paul}) as obtained from $\pi N$ 
scattering. Similar to previous studies \cite{WiA93,Wil96}, we 
do not determine the free parameters by  a {\it global} fit to all 
$NN$ channels in the $\Delta$ region. Instead, we concentrate on the 
most important $^1D_2$ channel, because it is the only partial wave 
which couples to a $N\Delta$-$S$ state (see Table~\ref{kap6_partial}) 
and thus where one expects the strongest $N\Delta$ interaction effects.

In order to distinguish the various cases, we introduce as nomenclature
``CC(approach, mesons)'', where for ``approach'' we consider a retarded one 
(``ret'') and different static ones (``stat'', ``stat1'', and ``stat2'').
The entry ``meson'' refers to the mesons included in the $N\Delta$ 
interaction, i.e., ``$\pi$'' means 
only pion exchange and ``$\pi$, $\rho$, $\alpha_{\Delta N\pi}$'' means 
$\pi$ and $\rho$ exchange including the additional coupling 
(\ref{kap3_lagrange2_rnd}) whose strength is controlled by 
$\alpha_{\Delta N\pi}$.
We begin with the discussion of the parameter choices for the retarded 
approach as listed in Table~\ref{kap6_tab1}. For ${\bar f}^0_{\Delta N\pi}$, 
${\bar f}^0_{\Delta N \rho}$, and the cutoff ${\bar \Lambda}_{\Delta N\pi}$, 
we have chosen the values of the full Bonn Potential, i.e.,
 \begin{eqnarray}\label{kap6_f_drn}
\frac{\left({\bar f}^0_{\Delta N\pi}\right)^2}{4\pi} &=& 0.224\,,\quad
\frac{\left({\bar f}^0_{\Delta N \rho}\right)^2}{4\pi} = 20.45\,,\quad
{\bar \Lambda}_{\Delta N\pi}(\vec{q}\,^2) = 1200 \, \mbox{MeV}
\,, \quad  {\bar n}_{\Delta\pi}=1\,.
\end{eqnarray}
It turned out that for a variety of different combinations of the remaining 
parameters  $\alpha_{\Delta N\rho}$ and ${\bar \Lambda}_{\Delta N \rho}$ a 
satisfactory 
description of the $^1D_2$ phase shift is possible. Thus we have restricted 
the choice of $\alpha_{\Delta N\rho}$ to the values $-1,\, 0,\, 1$. 
The resulting cutoff masses ${\bar \Lambda}_{\Delta N \rho}$ are listed in 
Table~\ref{kap6_tab1} as determined by a fit to the $^1D_2$ phase
shift at $T_{lab} \approx 500$ MeV. 
The value $\alpha_{\Delta N\rho}=1$ corresponds to the usual neglect of the 
additional coupling  (\ref{kap3_lagrange2_rnd}). On the other hand, the 
choice $\alpha_{\Delta N\rho}=0$ can be motivated by vector dominance insofar
as for this choice the $\rho {\bar N}\Delta$ vertex has the same
spin structure as the dominant $M1$ $\gamma {\bar N} \Delta$ coupling.

As next, we turn to the static case. We would like to remind the reader 
that in this case {\it all} diagrams except the contribution (c) of 
Fig.~\ref{abb_delta} are incorporated into the static part of $V^{0\,[2]}$,
where {\it all} corresponding coupling constants ($f^0_{\pi}$, $f^0_{\rho}$, 
$g^0_{\rho}$, ${\bar f}^0_{\Delta N\pi}$, ${\bar f}^0_{\Delta N\rho}$),
the parameter $\alpha_{\Delta N\rho}$ and {\it all} form factors $F_{\pi}$, 
${\bar F}_{\Delta N\pi}$ and ${\bar F}_{\Delta N\rho}$ are treated as 
free parameters. The resulting values are listed in Table~\ref{kap6_tab2}. 
In the simplest approach, CC$(\mbox{stat1},\pi)$, which is identical to the 
one used in \cite{WiA93}, we neglect the $\pi d$ channel and the 
$\rho$ exchange in $V^{0\,[2]}_{N\Delta}$ and $V^{0\,[2]}_{\Delta \Delta}$ 
completely, e.g.,
the $N \Delta$ interaction is solely given by $\pi$ exchange, where 
the cutoff $\Lambda_{\pi NN} = {\bar \Lambda}_{\Delta N\pi}$
is fitted to the $^1D_2$ partial wave at $T_{lab} \approx 500$ MeV. 
The $\pi N \Delta$ vertex $V^0_{\Delta \pi}$ is given by the 
parametrization (\ref{kap3_form_paul}) of Sauer {\it et al}.~\cite{PoS87}.
Similar to the retarded case, we use these parameters (\ref{kap3_form_paul})
also in the potential $V^{0\,\pi}_{\Delta \Delta}$ (diagram (f) of
Fig.~\ref{abb_delta}). 

The second approach CC(stat2, $\pi$) differs from CC(stat1, $\pi$) only 
with respect to the parametrization (\ref{kap3_form2}) instead of 
(\ref{kap3_form_paul}) for $V^0_{\Delta \pi}$, 
e.g., those values for $V^0_{\Delta \pi}$ are taken into account which are 
also present in our retarded approach.
The choice (\ref{kap3_form2}) is also used in our third case 
CC(stat, $\pi,\,\rho,\,0$). But in contrast to the previous cases, 
we incorporate in addition $\rho$ exchange in 
$V^{0\,[2]}_{N\Delta}$ and $V^{0\,[2]}_{\Delta \Delta}$ as well as the 
$\pi d$ channel. Furthermore, we set $\alpha_{\Delta N\rho} =0$.

\subsection{Results}\label{kap6_r}

\begin{figure}[btp]
\centerline{\psfig{figure=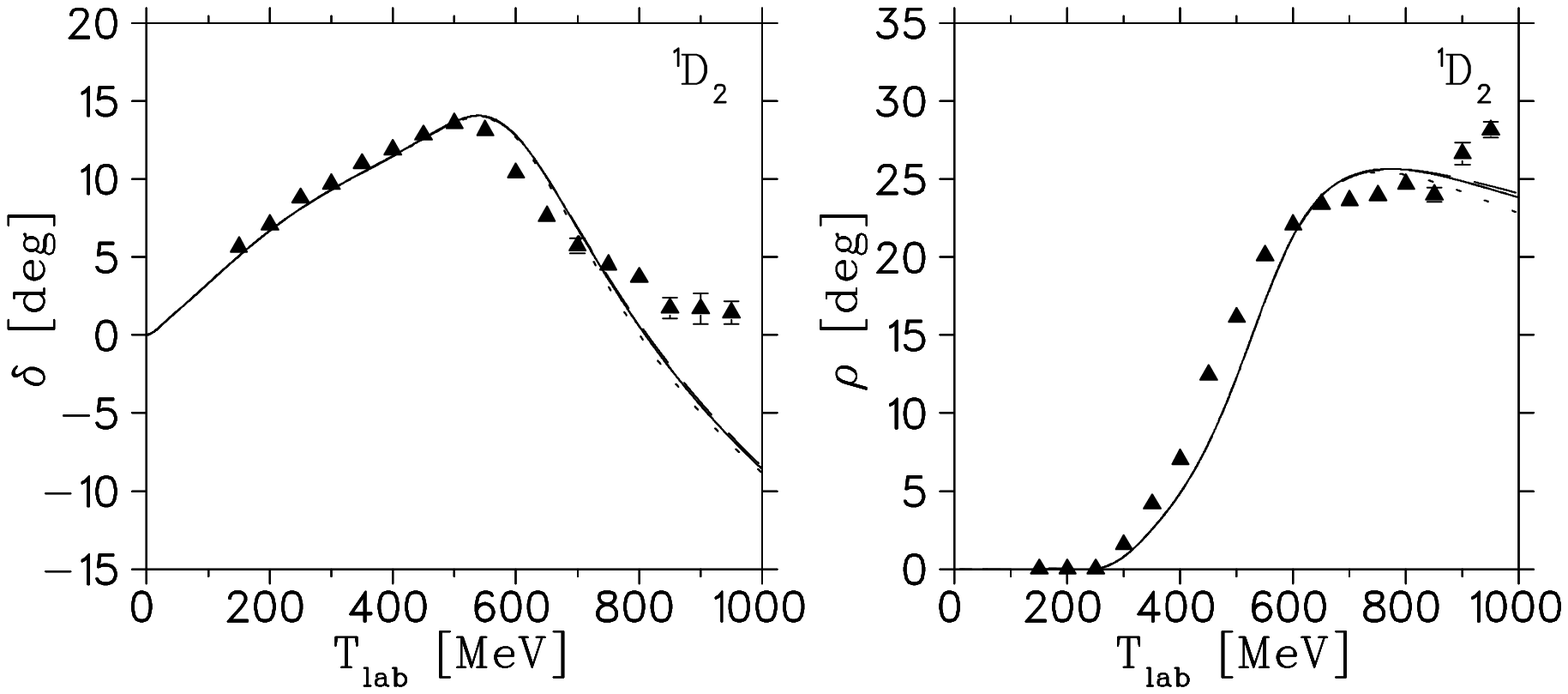,width=13cm,angle=0}}
\vspace{0.5cm}
\caption{Phase shift $\delta$ and inelasticity $\rho$ for the
$^1D_2$ channel in comparison with experiment (solution SM97 of
Arndt {\it et al.}~{\protect\cite{Arn98}})
for different retarded potential models
   (see Table {\protect \ref{kap6_tab1}}):
 full curve: CC(ret$,\pi,\rho,0)$, 
 dotted curve: CC(ret$,\pi,\rho,1)$, and
 dashed curve: CC(ret$,\pi,\rho,-1)$.}
\label{kap6_phasen_alpha1}
\end{figure}

\begin{figure}[pt]
\centerline{\psfig{figure=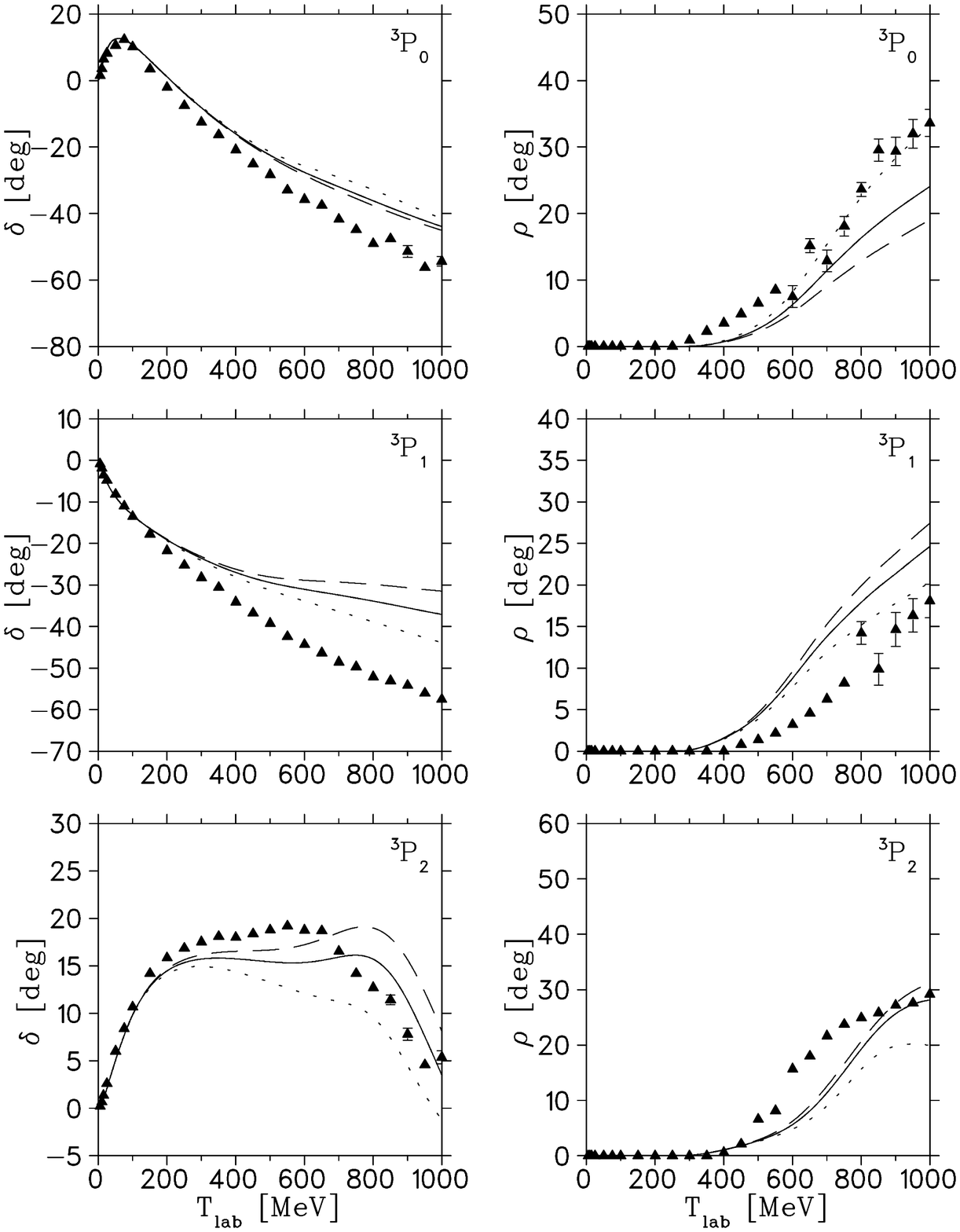,width=13cm,angle=0}}
\vspace{0.5cm}
\caption{Phase shift  $\delta$ and inelasticity $\rho$ for the 
$^3P_0$, $^3P_1$, and $^3P_2$ channels for different retarded potential 
models as in Fig.~\ref{kap6_phasen_alpha1}. Notation 
of the curves as in Fig.~\ref{kap6_phasen_alpha1} and experimental data: 
solution SM97 of Arndt {\it et al.}~{\protect\cite{Arn98}}.}
\label{kap6_phasen_alpha2}
\end{figure}

We will start the discussion with the $^1D_2$ partial wave. 
As already mentioned, all three retarded potential models are able to 
describe its phase shift and inelasticity equally well in the $\Delta$ 
region (see Fig.~\ref{kap6_phasen_alpha1}). This is of course not very
surprising because we had used this channel for fixing the free parameters. 
However, above $T_{lab} \geq 800$ MeV one notes a rather large discrepancy 
between theory and experiment. In view of the fact that the parameter 
$\alpha_{\Delta N \rho}$ could not be fixed uniquely, we have studied the 
influence of different choices for $\alpha_{\Delta N \rho}$ on the other 
partial waves. It turned out that the strongest dependence was found for 
the $P$ waves (see Fig.~\ref{kap6_phasen_alpha2}). However, it was not 
possible to determine an optimal value for $\alpha_{\Delta N\rho}$. While 
the $^3P_0$ inelasticity and the $^3P_1$ channel seem to favour 
$\alpha_{\Delta N\rho} = 1$, the value $\alpha_{\Delta N\rho}=-1$ leads to 
a slightly better fit of the $^3P_2$ channel, especially for energies 
$T_{lab}\leq 700$ MeV. But the overall description is still quite poor. 
On the other hand, this rather unsatisfactory situation is not very surprising 
since we have not fitted our parameters to all partial waves simultaneously. 

\begin{figure}[hpt]
\centerline{\psfig{figure=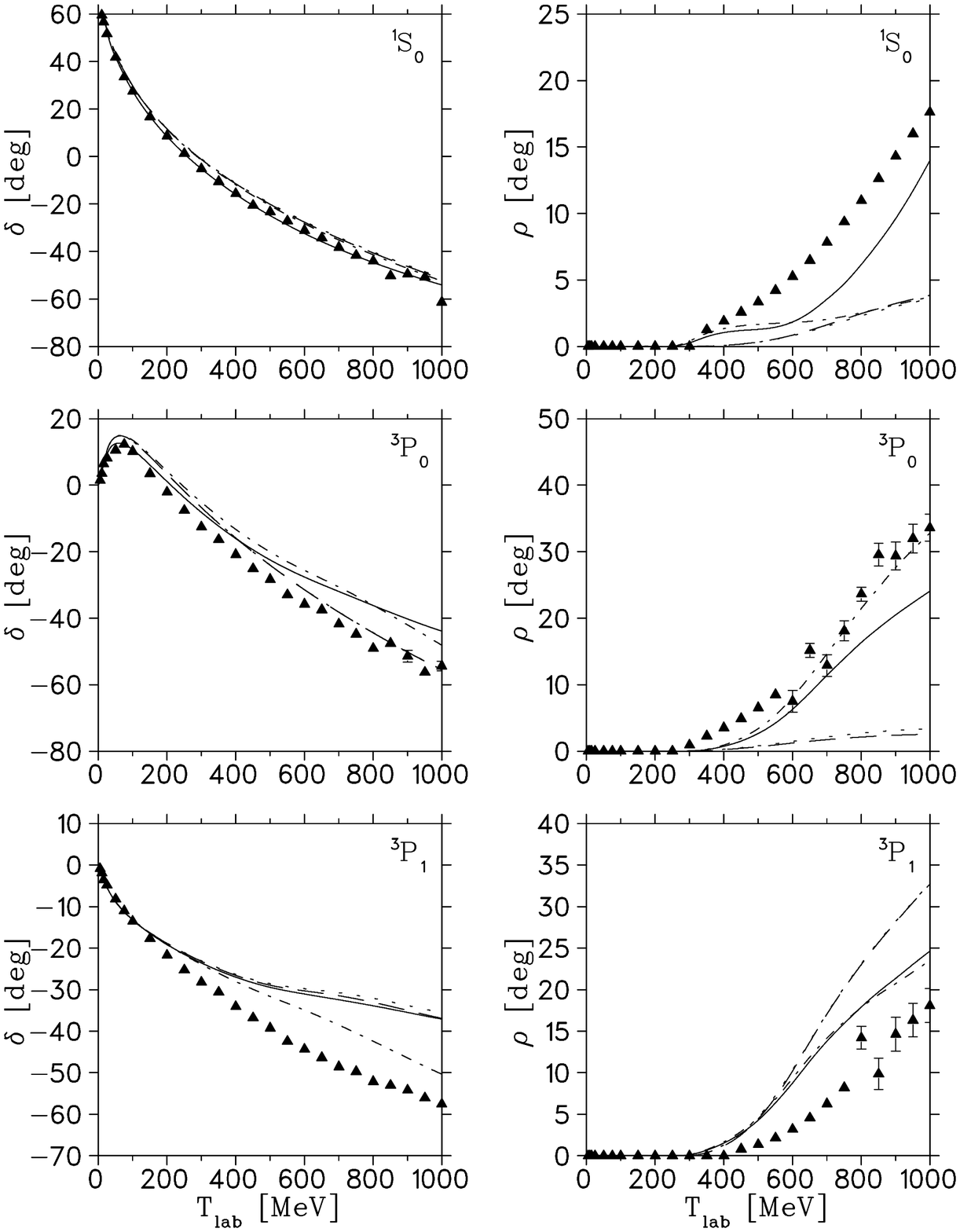,width=13cm,angle=0}}
\vspace{0.5cm}
\caption{Phase shift  $\delta$ and  inelasticity  $\rho$ for
the $^1S_0$ , $^3P_0$  and $^3P_1$ channels for different static 
and retarded approaches:
dotted curve: CC$(\mbox{stat1},\pi)$, dashed curve: CC$(\mbox{stat2},\pi)$,
dash-dotted curve: CC$(\mbox{stat},\pi,\rho,0)$,  
and full curve: CC(ret$,\pi,\rho,0)$. The experimental data 
represent solution SM97 of Arndt {\it et al.}~{\protect\cite{Arn98}}.}
\label{kap6_phasen1_vergleich}
\end{figure}

\begin{figure}[hpt]
\centerline{\psfig{figure=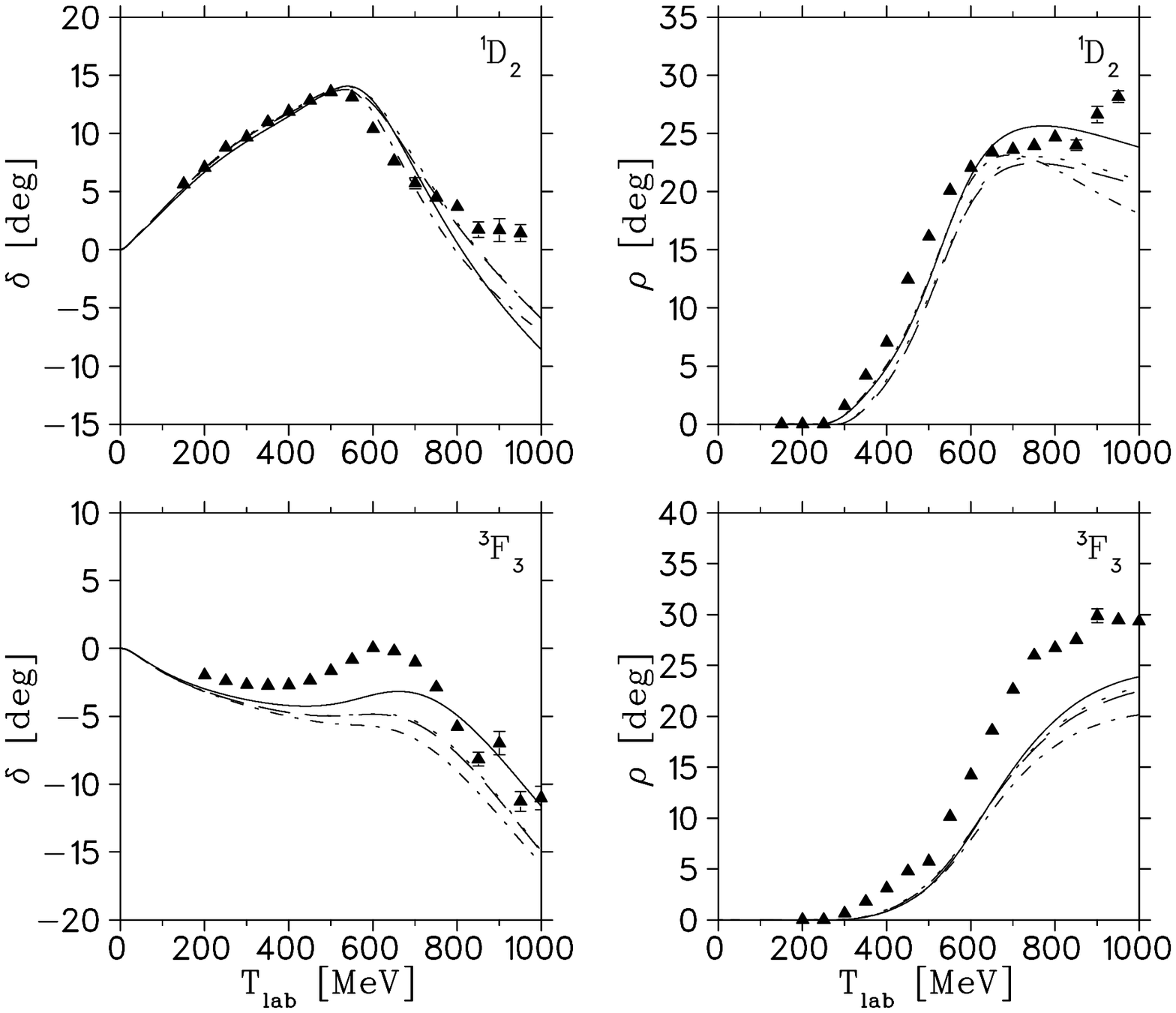,width=13cm,angle=0}}
\vspace{0.5cm}
\caption{Phase shift  $\delta$ and  inelasticity $\rho$ for the 
$^1D_2$  and  $^3F_3$ channels for different static 
and retarded approaches as in Fig.~\ref{kap6_phasen1_vergleich}. 
Notation  of the curves as in 
Fig.~\ref{kap6_phasen1_vergleich} and experimental data: 
solution SM97 of Arndt {\it et al.}~{\protect\cite{Arn98}}.}
\label{kap6_phasen2_vergleich}
\end{figure}

\begin{figure}[hpt]
\centerline{\psfig{figure=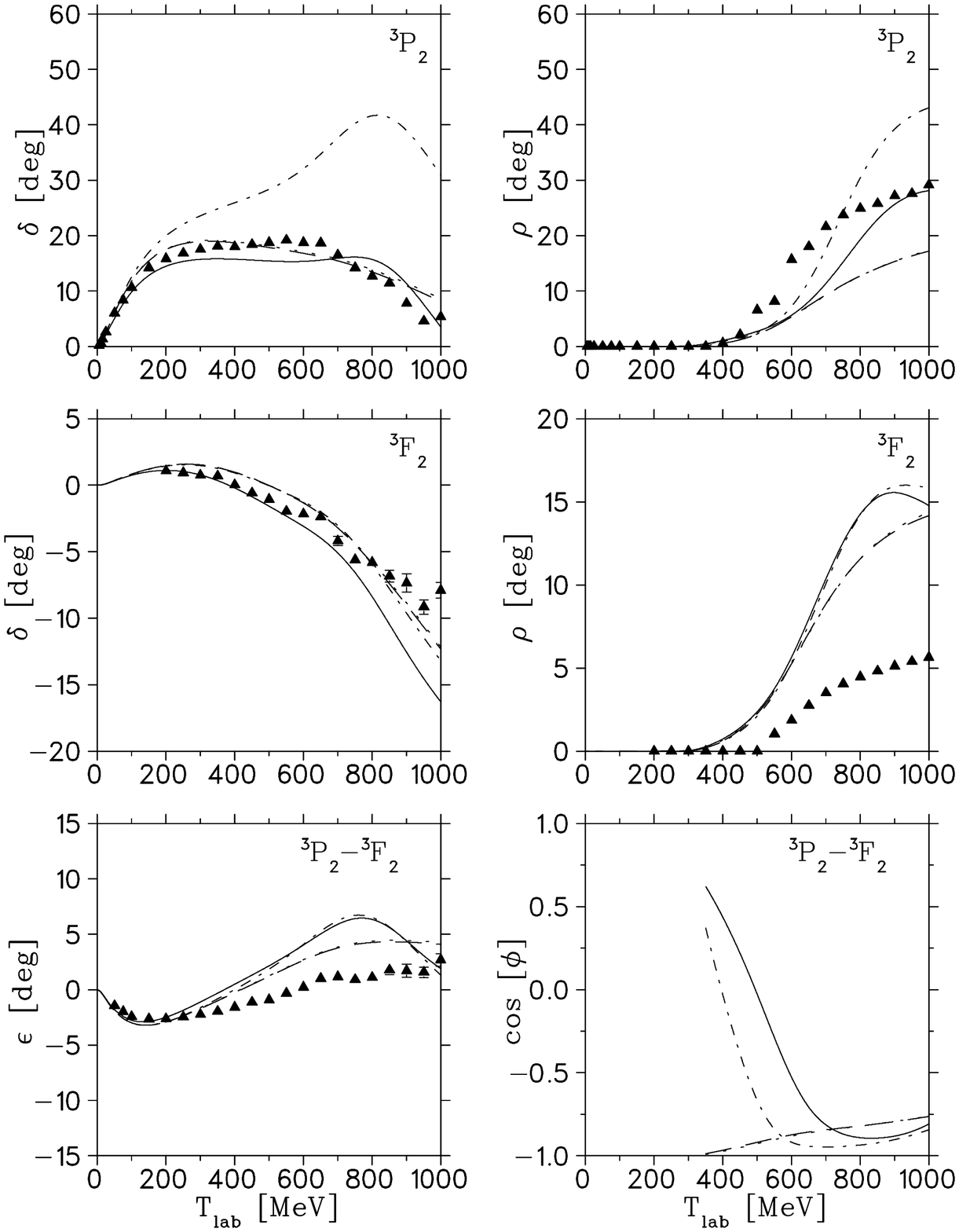,width=13cm,angle=0}}
\vspace{0.5cm}
\caption{Phase shifts, inelasticities and mixing angles for the
 $^3P_2$-$^3F_2$ channel for different static and retarded approaches 
as in Fig.~\ref{kap6_phasen1_vergleich}. Notation of the curves as in 
Fig.~{\protect\ref{kap6_phasen1_vergleich}} and experimental data: 
solution SM97 of Arndt {\it et al.}~{\protect\cite{Arn98}}.}
\label{kap6_phasen4_vergleich}
\end{figure}

Now we will turn to the discussion of the various static and retarded 
approaches for the hadronic interaction. We have chosen the model 
CC(ret, $\pi,\,\rho,\,0$)
as a starting point and compare it with the static approaches 
CC(stat1, $\pi$), CC(stat2, $\pi$), and CC(stat, $\pi,\,\rho,\,0$) 
for all $(t=1)$ partial waves with total angular momentum $j \leq3$ as shown
in Figs.~\ref{kap6_phasen1_vergleich} through \ref{kap6_phasen4_vergleich}. 
With respect to the $^1D_2$ phase shift, all four models yield more or 
less similar results. The inelasticity in CC(stat1, $\pi$) and 
CC(stat2, $\pi$), however, is somewhat smaller than in the other two models. 
In accordance with \cite{TaO87}, this behaviour can be traced back to the 
neglect of the $\pi d$ channel in CC(stat1, $\pi$) and  CC(stat2, $\pi$). 
For the other partial waves, the comparison with experimental data does 
not give a uniform picture. For example, the calculation 
CC(stat, $\pi,\,\rho,\,0$) yields the comparably best, but still not 
satisfactory description of the $^3P_1$ phase shift and the 
$^3P_0$ inelasticity. On the other hand, the theoretical $^3P_2$ phase 
shift is largely at variance with the data, whereas CC(stat1, $\pi$) and 
CC(stat2, $\pi$) result in a rather good description. But in the
$^1S_0$ and $^3F_3$ channels the retarded calculation 
CC(ret, $\pi,\,\rho,\,0$) is favoured. 

In summary, one finds that our fit procedure leads to a satisfactory 
agreement with the experimental data for the $^1D_2$ channel, which is 
the most important one in the $\Delta$ resonance region. The overall 
description of the other channels is fairly well but needs further 
improvements. As a first step one would need to construct a potential model 
(static or retarded) in which {\it all} parameters are 
fitted simultaneously to {\it all} partial waves for energies up to about
$T_{lab} =1$ GeV. This quite involved task will be one of our future projects.

Another remark is in order with respect to the question whether the static 
or the retarded interaction gives a better description. Although in principle, 
from the basic physical ideas, a retarded interaction is more appropriate, our 
results for the pure hadronic reactions do not allow a clear answer. 
To this end, one has to consider e.m.\ reactions where 
first results~\cite{Sc198,Sc298} indicate a preference for the retarded 
interaction. This will be the subject of forthcoming papers. 

\subsection{Comparison with other approaches}\label{kap6_w}

In the coupled $NN$-$N \Delta$ approach of Leidemann 
{\it et al}.~\cite{LeA87}, the static Reid soft core and Argonne potentials 
have been used as starting point for the $NN$ interaction, 
whereas the $\pi d$ channel has been neglected. Since the 
numerical calculations have been performed in r-space, nonlocalities,
which are, for example, present in the $N \Delta$ propagation, could not 
be taken into account exactly. Therefore, additional approximations in the
box-renormalization procedure had been used which lead to a much more flexible
treatment by introducing one additional free parameter for each $NN$ 
partial wave.
Thus, compared with our results, a better description of $NN$ scattering 
data was achieved because of this additional phenomenological ingredient. 
With respect to the static framework, we further would like to mention 
the work of Wilbois~\cite{Wil96}. Among other things, he compared the 
results of a coupled channel approach for different static NN potentials 
(Bonn-OBEPR, Bonn-OBEPQ, Paris and Nijmegen). It turned out that the 
$NN$ phase shifts and inelasticities are rather sensitive to the choice 
of the underlying $NN$ potential.  No $NN$ potential, incorporated into 
the CC approach, has been able to produce a quantitative description of 
all $NN$ phases and inelasticities in the $\Delta$ resonance region.

Retarded approaches based on three-body theory \cite{KlS80a,TaO88}
have already been discussed in Sect.~\ref{introduction}. Here, we want 
to consider in some more detail the work of Bulla {\it et al}.\ and Elster 
{\it et al}.\ \cite{ElH88,BuS92}. 
The coupled channel approach of \cite{BuS92}, which includes an explicit 
$\pi {\bar N}$ vertex, differs at least in three important aspects compared 
to the present approach: (i) The $\pi d$ channel is neglected. 
(ii) Retardation is treated only for pion exchange. For the cutoff mass 
of the corresponding $\pi {\bar N}$ vertex form factor, a very small value
of 443 MeV has been used in dipole parametrization as obtained
from a fit of $\pi N$ scattering data in the $P_{11}$ channel, whereas the
value of 1700 MeV in the Elster potential results from a fit to 
$NN$ scattering data.
(iii) The static Paris potential is used as the underlying $NN$ interaction. 
Therefore, because of the renormalization procedure (\ref{kap3_sau3}), 
the effective retarded OPE potential is considerably weakened.
In view of the points (ii) and (iii), it is not very surprising that Bulla 
{\it et al}.\ found only a small influence from retardation, in 
fact much smaller than in our approach. Moreover,  
due to the mixture of retarded and static frameworks 
it is rather questionable to apply this approach to electromagnetic
reactions on the deuteron. For example, due to the existence of two different
$\pi$ exchange mechanisms in the Paris potential and the retarded interaction,
two different types of MECs have to be constructed in order to guarantee
gauge invariance - which is rather artificial.

In the work of Elster {\it et al}.\ \cite{ElH88},
who have extended the Bonn boson exchange potential into the region above 
pion threshold by the inclusion of the $\Delta$ isobar, the interactions 
are treated in a retarded approach including appropriate nucleon and 
$\Delta$ dressing. 
Elster {\it et al}.\ start within a Lee model framework
whereas our retarded interactions are  derived 
within time-ordered perturbation theory. In 
contrast to our model, the $\pi d$ channel 
as well as the ${\bar N} \Delta \leftrightarrow {\bar N} \Delta$ 
transition potential (diagram (c) and (f) in Fig.~\ref{abb_delta}) are 
neglected, so that unitarity is violated above $\pi$ threshold. On the 
other hand, the $\Delta$ is treated relativistically whereas we describe it 
nonrelativistically. Moreover, $\Delta\Delta$ components are taken 
into account in \cite{ElH88}.  Concerning the pure nucleonic sector, 
however, both approaches yield exactly the same interaction.

\section{Summary and Outlook}\label{summary}

In this paper, we have presented a hadronic interaction model
which is suited for the study of electromagnetic and hadronic reactions 
in the two-nucleon sector for exitation energies up to about 500 MeV, 
for example, $NN$ scattering or electromagnetic deuteron break-up, in which 
not more 
than one pion is created or absorbed. This model respects in particular 
three-body unitarity. It is based on a previously developed model 
of Sauer and collaborators \cite{BuS92,Sau86,PoS87} which allows
the explicit consideration of retardation in the two-body meson-exchange 
operators within a $NN$-$N\Delta$ coupled channel approach using time-ordered 
perturbation theory. 

Since retarded interactions are not hermitean, we have generated the 
retarded one-boson exchange mechanisms by considering explicitly 
meson-nucleon  and $\pi N \Delta$ vertices. Therefore, additional 
mesonic degrees of freedom besides the baryonic components had to be 
included explicitly into the Hilbert space of a two-baryon system. 
In the present work, we have restricted ourselves to the 
one-meson-approximation, which means that we allow only configurations 
with one 
meson present besides the baryons. As mesons we have taken into account 
explicitly $\pi,\, \rho,\, \omega,\, \sigma,\, \delta$, and $\eta$. In 
order to satisfy two- and three-body unitarity, we have 
incorporated the $\pi d$ channel as well as intermediate pion-nucleon loops. 
In view of the latter mechanism, a distinction between bare and 
physical nucleons was necessary in order to avoid inconsistencies. 

In our explicit realization within a field-theoretical framework, we have 
used the realistic retarded potential model of Elster 
{\it et al.}~\cite{Els86,ElF88} as input for the basic nucleon-nucleon 
interaction in pure nucleonic space. This potential had to be renormalized 
because of the additional d.o.f., which lead to further contributions from 
$\bar N\Delta$  and $\pi d$ states. Because of some necessary approximations 
in our numerical evaluations, unitarity is not completely obeyed, which 
fact, however, we do not consider very critical. The 
free parameters of our model have been fixed
by fitting $\pi N$ scattering in the $P_{33}$ channel and $NN$ scattering
in the $^1D_2$ channel. For practical reasons, no global fit of the
parameters to {\it  all} relevant $NN$  or $\pi N$ channels has been performed.
Therefore, the overall description ot the $NN$ phase shifts and inelasticities
is fairly well but needs some further improvement in the future.
Moreover, due to our fit procedure, there remain some ambiguities because
not all parameters could be fitted uniquely. For example, besides the usual 
$\rho \bar{N} \Delta$ interaction density (\ref{kap3_lagrange1_rnd}) 
we have incorporated an additional alternative (\ref{kap3_lagrange2_rnd}) 
into the $\rho$ exchange of the $\bar{N} \bar{N}\rightarrow 
\bar{N} \Delta$ transition, which has, at least to our
knowledge, not yet been discussed in the literature. The corresponding
coupling constant could not be fixed unambiguously by studying the
$^1D_2$ partial wave alone. However, different choices of this
coupling constant lead to rather drastic changes in the $P$ wave
phase shifts in the $\Delta$ region. 
For an improved description one would need to construct from scratch 
a hadronic interaction model which incorporates from the beginning nucleon, 
meson, and $\Delta$ degrees of freedom and whose open parameters are fitted 
to the phase shifts and inelasticities of {\it all relevant} $NN$ scattering 
partial waves for $T_{lab}$ energies up to about 1 GeV. 

Another considerable improvement would be the implementation of the 
convolution approach of Kvinikhidze and Blankleider  
\cite{KvB93a,KvB93b,BlK94} into our model. This would allow one to abandon 
the one-meson-approximation which causes several pathologies. From a 
conceptual point of view, this problem deserves further attention
because it deals with the fundamental question of deriving off-shell
properties  of particles in a many-body system (for example the dressing) 
from those of a single-particle system without creating inconsistencies.
 
In conclusion, we believe that the present model is realistic enough for 
the study of electromagnetic reactions like photo- and electrodisintegration 
of the deuteron as well as pion production. The predictions for the 
observables of such reactions will be presented in forthcoming papers.

\renewcommand{\theequation}{A\arabic{equation}}
\setcounter{equation}{0}
\section*{Appendix A: The physical nucleon}
\label{appA}

Here we will consider briefly a ``physical'' nucleon state $\ket{N}$.
It is straightforward to show that the physical nucleon state can be written as
\begin{equation}\label{kap3_phys_np2}
\ket{N(\vec{p}\,)} = N_{[1]}^{-1}(p)\, \left( 1 +
g_0( e_N(p)) v^0_{X{\bar N}} \right) \ket{{\bar N}(\vec{p}\,)}\,,
\end{equation}
where the free propagator $g_0(z)$ is defined by 
\begin{equation}\label{kap3_g0_1}
g_0(z) =(z-h_0)^{-1}\,,
\end{equation}
and $v^0_{X{\bar N}}$ is the only interaction term describing the emission 
of a meson, since we do not consider a diagonal meson-nucleon 
interaction in ${\cal H}^{[1]}_{X}$, i.e., $v^0_{XX}$ vanishes.  
Note that a $\Delta$ will not appear because of isospin conservation and the 
one-meson-approximation. For the renormalization constant one finds 
\begin{equation}\label{kap3_nphys_norm}
N_{[1]}(p) = \sqrt{1 + v^2_{[1]}(p)} \,,
\end{equation}
where $v^2_{[1]}(p)$ is defined by 
\begin{equation}
\bra{{\bar N}(\vec{p}^{\,\prime}\,)}
v^0_{{\bar N}X} g_0( e_N(p^{\,\prime}))g_0( e_N(p)) 
v^0_{X{\bar N}} \ket{{\bar N}(\vec{p}\,)}= v^2_{[1]}(p)
\delta(\vec{p}^{\,\prime}-\vec{p}\,)\,.
\end{equation}
Furthermore, for the counter term one obtains the identity 
\begin{equation}\label{kap3_counter1}
\bra{{\bar N}(\vec{p}^{\,\prime})}
v^{[c]}_{{\bar N}{\bar N}} \ket{{\bar N}(\vec{p}\,)}= 
- \bra{{\bar N}(\vec{p}\,)} 
v^0_{{\bar N}X} \,(e_N(p) - 
 h_{0,\,XX})^{-1}  v^0_{X{\bar N}}
\ket{{\bar N}(\vec{p}\,)}  
\, \delta(\vec{p}^{\, \prime} -\vec{p}\,)\,,
\end{equation}
yielding the following compact expression 
\begin{equation}\label{kap3_nphys_counter}
 v^{[c]}_{{\bar N}{\bar N}} = 
-\int dz\,\delta(z-H_0)\,v^0_{\bar N X} \,g_0(z)\,v^0_{X\bar N}\,.
\end{equation}
In summary, it is obvious that a distinction between bare and physical 
nucleons is necessary due to the occurrence of the interaction 
$v^0_{\bar N X}$ which generates intermediate meson-nucleon loops.

\renewcommand{\theequation}{B\arabic{equation}}
\setcounter{equation}{0}
\section*{Appendix B: Derivation of the $NN$ scattering $T$-matrix}
\label{appB}

\begin{figure}[hp]
\centerline{\psfig{figure=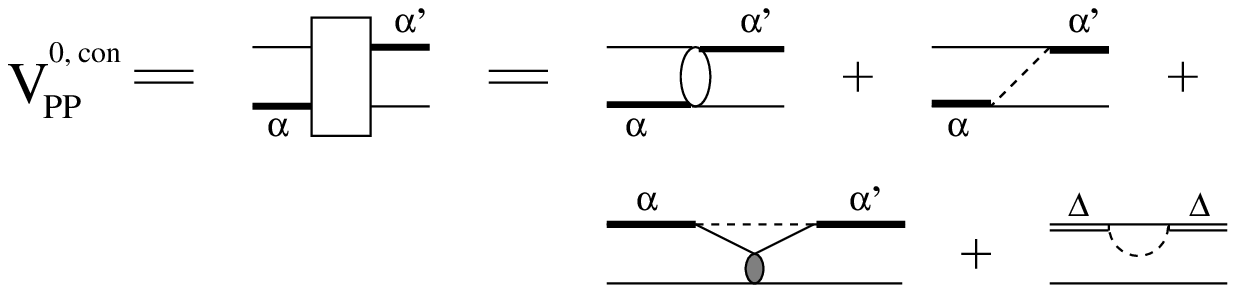,width=12cm,angle=0}}
\vspace{0.5cm}
\caption{Diagrammatic representation of the connected driving term 
$V^{0,\,con}_{PP}(z)$. The greek letters $\alpha$ and $\alpha'$ label 
either a bare nucleon ${\bar N}$ or a $\Delta$.}
\label{fig_V_0_con}
\end{figure}

\begin{figure}[hp]
\centerline{\psfig{figure=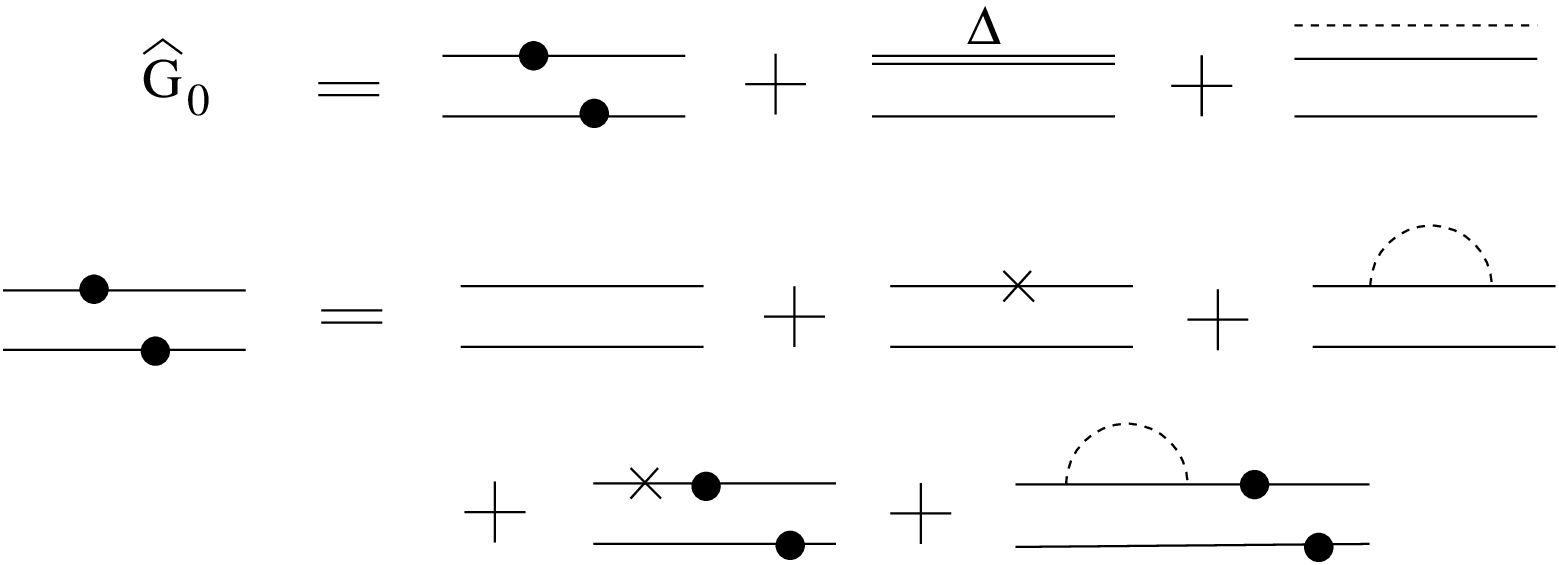,width=12cm,angle=0}}
\vspace{0.5cm}
\caption{Diagrammatic representation of the dressed propagator 
$\widehat G_0(z)$. The cross represents the one-nucleon counter term 
$v^{[c]}$.}
\label{fig_G_0_hat}
\end{figure}

\begin{figure}[hp]
\vspace{0.5cm}
\centerline{\psfig{figure=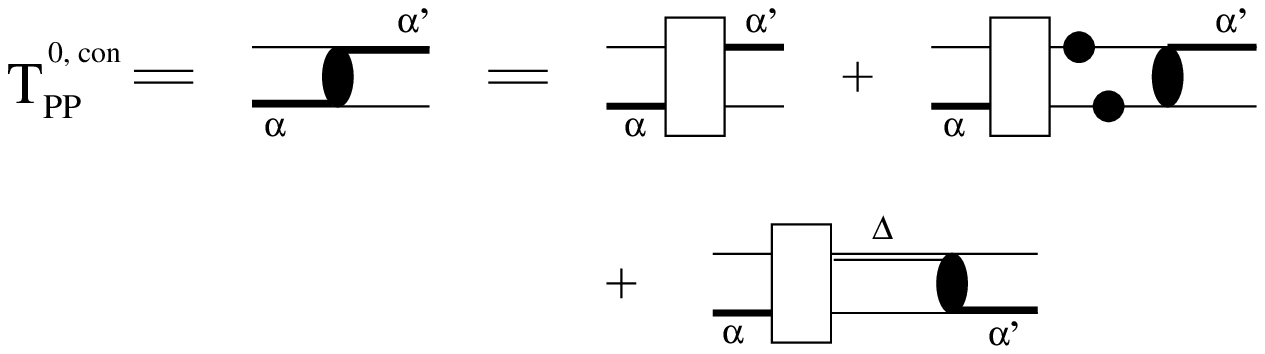,width=12cm,angle=0}}
\vspace{0.5cm}
\caption{Graphical representation of the scattering amplitude 
$T^{0,\, con}_{PP}(z)$. The greek letters $\alpha$ and $\alpha'$ label 
either a bare nucleon ${\bar N}$ or a $\Delta$. The driving term 
$V^{0,\,con}_{PP}(z)$ is shown in Fig.~\protect\ref{fig_V_0_con}.}
\label{fig_T_0_con}
\end{figure}

In this appendix we will briefly sketch the derivation of 
Eq.~(\ref{kap3_unn2_mod}). With the help of projection operators one can 
bring (\ref{kap3_nn_s_mod}) into the equivalent form   
\begin{eqnarray}
\ket{NN;\,\vec{p}, \alpha}^{(\pm)}
 &=& N_{[2]}^{-1}(\vec{p}\,)  \Big(  1+
 G_{0}(E^{NN}_{p} \pm i\epsilon)  
 T^0_{P\bar N}( E^{NN}_{p} \pm i\epsilon) + G_{0}(E^{NN}_{p} \pm i\epsilon)  
 T^0_{X\bar N}( E^{NN}_{p} \pm i\epsilon) \Big)
 \ket{{\bar N}{\bar N};\,\vec{p}, \alpha} \, ,\label{kap3_nn_s2}
\end{eqnarray}
where  $T^0_{PP}$ and $T^0_{XP}$ are given by 
(note $P=P_{\bar N}+P_{\Delta}$) 
\begin{eqnarray}
 T^0_{PP}(z) &=& \bar V^{0}_{PP}(z) + 
\bar V^{0}_{PP}(z)
  G_0(z) T^0_{PP}(z) \,,  \label{kap3_tpp2_mod} \\  
T^0_{XP}(z) &=& G_0^{-1}(z)G^X(z) V_{XP}^0
 \Big( 1+G_0(z) T^0_{PP}(z) \Big) \, .
 \label{kap3_tqp2_mod} 
\end{eqnarray}
In these relations, $\bar V^{0}_{PP}$ is defined  by
\begin{equation}\label{kap3_hatv0_mod}
 \bar V^{0}_{PP}(z) = V_{PP}^0 + V_{PX}^0 
 G^X(z) V_{XP}^0 \,, 
\end{equation}
and $G^X(z)$ in (\ref{kap3_vqq3_mod}).
The driving term $\bar V^{0}_{PP}$ in (\ref{kap3_hatv0_mod}) can now be 
split into a connected (``con'') and a disconnected (``dis'') part
\begin{equation}\label{kap3_dis_con}
  \bar V^{0}_{PP}(z) = V^{0,\,dis}_{PP}(z) + V^{0,\,con}_{PP}(z) \,, 
\end{equation}
where, by definition, $V^{0,\,dis}_{PP}$ contains only those parts of the 
driving term $\bar V^{0}_{PP}$ which do not describe an 
interaction between the two baryons except for the $\pi N$ loop 
contributions to the $\Delta$ self energy, which we have incorporated into 
$V^{0,\,con}_{PP}(z)$, containing otherwise the genuine baryon-baryon 
interaction. In detail one finds from
(\ref{kap3_hatv0_mod}) and (\ref{kap3_vqq3_mod}) with (\ref{kap3_vnn})
\begin{eqnarray}
V^{0,\,dis}_{PP}(z) &=& V^{[c]}_{{\bar N}{\bar N}}
+ \left[V^0_{{\bar N}X} G_0(z) V^0_{X{\bar N}} \right]_{dis}\,,
\label{kap3_vdiss_mod}\\
V^{0,\,con}_{PP}(z) &=& V^{0\,[2]}_{PP} + 
 \left[V^0_{PX} G_0(z) V^0_{XP} \right]_{con}
 + V^0_{PX} G_0(z)T^X(z)G_0(z) V^0_{XP}
 + \left[V^0_{\Delta X} G_0(z) V^0_{X\Delta}\right]_{dis}
\,,\label{kap3_vcon_mod}
\end{eqnarray}
where we have defined 
\begin{eqnarray}
\left[V^0_{\alpha X} G_0(z) V^0_{X\alpha } \right]_{dis}
&=&  \sum_{i=1,2}    
v^0_{\alpha X}(i) G_0(z) v^0_{X\alpha }(i)\,, \quad 
\alpha \in \{\bar N,\,\Delta\}\,,
\label{kap2_v_cd2}\\
\left[V^0_{\alpha X} G_0(z) V^0_{X\beta } \right]_{con}
&= & \sum_{i,j=1,2;\,j \neq i}
v^0_{\alpha X}(i)  G_0(z)  v^0_{X\beta }(j)\,, \quad 
\alpha,\,\beta \in \{\bar N,\,\Delta\}\,.
\label{kap3_obe_mod}
\end{eqnarray}
Thus the connected part $V^{0,\,con}_{PP}(z)$, which is shown in 
Fig.~\ref{fig_V_0_con}, contains besides $V^{0\,[2]}_{PP}$ 
a retarded one-boson exchange interaction, the coupling to 
the $\pi NN$ channel, and, furthermore, also a 
disconnected part, namely, the already mentioned $\pi N$ loop contributions 
to the $\Delta$ self energy in the last term of (\ref{kap3_vcon_mod}). 
The separation (\ref{kap3_dis_con}) allows us 
to represent the total amplitude by a ``disconnected'' 
$T^{0,\,dis}_{PP}(z)$ and a ``connected'' $T^{0,\, con}(z)$ amplitude 
\begin{equation}\label{kap3_t_z_mod}
T^0_{PP}(z) = T^{0,\, dis}_{PP}(z) + 
\left( 1 + T^{0,\, dis}_{PP}(z) G_0(z) \right)
 T^{0,\, con}_{PP}(z) \left( 1+  G_0(z) T^{0,\, dis}_{PP}(z) \right) \,,
\end{equation}
where
\begin{eqnarray}
T^{0,\,dis}_{PP}(z) &= &V^{0,\,dis}_{PP}(z)  + V^{0,\,dis}_{PP}(z)  
G_0(z) T^{0,\,dis}_{PP}(z)\,, 
\label{kap3_tdiss_1}\\
T^{0,\, con}_{PP}(z) &=& V^{0,\, con}_{PP}(z) + 
 V^{0,\, con}_{PP}(z) {\widehat G}_0(z) T^{0,\, con}_{PP}(z)\,,
\label{kap3_tcon_mod}
\end{eqnarray}
with the dressed propagator 
\begin{eqnarray}
{\widehat G}_0(z) &=& (z-H_0 - V^{0,\,dis}_{PP}(z))^{-1}
\nonumber \\
&=& G_0(z) + G_0(z) T^{0,\,dis}_{PP}(z) G_0(z) \, . \label{kap3_gquer1}
\end{eqnarray}
A graphical representation of the dressed propagator in the form 
${\widehat G}_0(z)=G_0(z) + G_0(z) V^{0,\,dis}_{PP}(z) \widehat G_0(z)$ 
is displayed in Fig.~\ref{fig_G_0_hat} and of the connected scattering 
amplitude $T^{0,\, con}_{PP}(z)$ in Fig.~\ref{fig_T_0_con}. 

We are now in the position to rewrite (\ref{kap3_nn_s_mod}) 
as follows
\begin{eqnarray}
\ket{NN;\,\vec{p}, \alpha}^{(\pm)}
 &=& N_{[2]}^{-1}(\vec{p}\,) 
 \left( 1+ G^X(E^{NN}_{p} \pm i\epsilon) V^0_{XP} \right) 
 \left(1+ {\widehat G}_0(E^{NN}_{p} \pm i\epsilon) T^{0,\, con}_{PP}
 (E^{NN}_{p} \pm i\epsilon) \right)  \nonumber \\ 
 & & \times
 \left(1+ G_0(E^{NN}_{p} \pm i\epsilon)  T^{0,\, dis}_{PP}(E^{NN}_{p} \pm i\epsilon)
 \right) \ket{{\bar N}{\bar N};\,\vec{p}, \alpha}\, .
\label{kap3_nn_s2_mod}
\end{eqnarray}
In order to determine the counter\-term $V^{[c]}_{{\bar N}{\bar N}}$ in the 
two-nucleon sector, we will consider the special case 
of two {\it noninteracting} physical nucleons denoted by 
$\ket{NN;\,\vec{p}, \alpha}$. Using the decomposition
\begin{equation}\label{kap3_nnphys}
\ket{NN;\,\vec{p}, \alpha} =  P_{\bar N} \ket{NN;\,\vec{p},\alpha} + 
P_X \ket{NN;\,\vec{p}, \alpha}\, 
\end{equation}
into nucleonic and mesonic components, one obtains, for example, for 
$P_{\bar N} \ket{NN;\,\vec{p},\alpha}$ a Schr\"odinger equation
\begin{equation}\label{kap3_nn1_mod}
\left( H_0 + V^{0,\,dis}_{PP}(E^{NN}_{p}) \right)     P_{\bar N} 
 \ket{NN;\,\vec{p},\alpha} =
 E^{NN}_{p}  \ket{NN;\,\vec{p},\alpha} \, .
\end{equation}
In view of $P_{\bar N}\ket{NN} \sim \ket{{\bar N}{\bar N}}$
and  the trivial identity $H_0 P_{\bar N} \ket{NN(\vec{p},\alpha)}
 = E^{NN}_{p}  P_{\bar N} \ket{NN(\vec{p},\alpha)}$, 
 one finds for the counter\-term 
(note the analogy to the one-nucleon sector)
\begin{equation}\label{kap3_nnphys_counter}
 V^{[c]}_{{\bar N}{\bar N}} = - \int dz\,\delta(z-H_0)\,
\left[V^0_{{\bar N}X}\, G_0(z)\, V^0_{X{\bar N}} \right]_{dis}\,.
\end{equation}

It should be emphasized that, due to the one-meson-approximation, one 
has to face the pathological situation that a dressing of both nucleons 
at the same time is not possible. Therefore, $\ket{NN;\,\vec{p}, \alpha}$
is {\it not} the direct product of two physical nucleon states of the 
one-nucleon sector, i.e., 
\begin{equation}\label{kap3_nn_n_dp}
\ket{NN(\vec{p}\,)} \not=   \ket{N(\vec{p}\,)}
 \otimes \ket{N(-\vec{p}\,)} \,\, .
\end{equation}
A way out of this well known  problem \cite{SaS85} could  be the 
``convolution approach'' of Kvinikhidze and Blankleider
\cite{KvB93a,KvB93b,BlK94}, which for technical reasons has not yet been 
implemented in our model.

With the help of the two-body renormalization operator 
${\widehat Z}_{[2]}(z)$ as defined in (\ref{kap3_renorm_mat_z2_2}), 
one can rewrite the dressed propagator ${\widehat G}_0(z)$ according to 
\begin{eqnarray}
{\widehat G}_0(z) 
&=&  G_0(z) \, {\widehat Z}^{-2}_{[2]}(z)\, ,
\label{kap3_renorm_mat_nenner_3}
\end{eqnarray}
and one obtains from (\ref{kap3_gquer1}) the identity 
\begin{equation}\label{kap3_z_mod}
1+ G_0(z)  T^{0,\, dis}_{PP}(z) =  {\widehat Z}^{-2}_{[2]}(z) \,, 
\end{equation}
which allows us to determine the renormalization constant 
$N_{[2]}(p)$ straightforwardly. From the normalization condition
\begin{equation}\label{kap3_nn_norm_mod}
\braket{NN;\,\vec{p}^{\,\prime}, \alpha^{\prime}}{NN;\,\vec{p}, \alpha}
= \delta_{\alpha^{\prime} \alpha}\delta ( \vec{p}^{\,\prime} - \vec{p} \,)\,, 
\end{equation}
one finds 
\begin{equation}\label{kap3_norm_3_mod}
 \bra{{\bar N}{\bar N};\,\vec{p}^{\,\prime}, \alpha}
({\widehat Z}_{[2]}^{os})^{-1}
 \ket{{\bar N}{\bar N};\,\vec{p}, \alpha}
 =  N_{[2]}(p) \delta_{\alpha^{\prime} \alpha} 
\delta ( \vec{p}^{\,\prime} - \vec{p}\,) 
\,.
\end{equation}
Introducing the ``renormalized'' interaction from 
(\ref{kap3_renorm_mat2_5_mod}), one obtains from 
 (\ref{kap3_tcon_mod}) with the help of 
(\ref{kap3_renorm_mat_nenner_3}) and (\ref{kap3_renorm_mat_r1}), 
 for the renormalized amplitude  
$T^{con}_{PP}(z)$, defined as
\begin{equation}
T^{con}_{PP}(z)={\widehat Z}^{-1}_{[2]}(z)\,T^{0,\,con}_{PP}(z)\,,
 {\widehat Z}^{-1}_{[2]}(z) \,
\end{equation}
the Lippmann-Schwinger equation in (\ref{kap3_tpp3_mod}), which is very 
useful for practical evaluations. In fact, considering $NN$ scattering 
for which the $T$ matrix is given by the onshell matrix element of $T^0$
\begin{eqnarray}
 N_{[2]}^{-2}(p)\,
 \bra{{\bar N}{\bar N};\,\vec{p}^{\,\prime}, \alpha'}
 T^0(E^{NN}_{p}+i\epsilon)
 \ket{{\bar N}{\bar N};\,\vec{p}, \alpha}\,,
 \label{kap3_s1_mod}
\end{eqnarray}
where $|\vec{p}^{\,\prime}| = |\vec{p}\,|=p$ has been used, one can rewrite  
the total amplitude $T^0_{PP}$ (\ref{kap3_t_z_mod}) as 
\begin{equation}\label{kap3_ton_mod_mod}
T^0_{PP}(z) = T^{0,\, dis}_{PP}(z) +
 {\widehat Z}^{-1}_{[2]}(z)\,T^{con}_{PP}(z)\,
 {\widehat Z}^{-1}_{[2]}(z) \,.
\end{equation}
With (\ref{kap3_norm_3_mod}) and the relation 
\begin{eqnarray} 
T^{0\,\,dis}_{PP}(z)
 \ket{{\bar N}{\bar N};\, \vec{p},\alpha}
&=& \left( {\widehat Z}^{-2}_{[2]}(z)-1 \right) G_0^{-1}(z)
\ket{{\bar N}{\bar N};\, \vec{p},\alpha} \,,
\label{kap3_mod_um}
\end{eqnarray}
which vanishes for $z = E^{NN}_{p} \pm i\epsilon$, one finds for 
the matrix element in Eq.\  (\ref{kap3_s1_mod}) the desired expression 
of (\ref{kap3_unn2_mod}).

\begin{table}
\caption{Meson parameters of $V^{Elster}$. The parameters for the
 $\sigma$ meson apply only to the $(t=1)$-$NN$ potential.
 For $t=0$, we use $g_{\sigma}^2/4\pi= 9.4050$, $m_{\sigma} = 580$ 
 MeV, $\Lambda_{\sigma}= 2300$ MeV,  $n_{\sigma} = 1$. 
}
\begin{center}
\begin{tabular}{ccccc}
$x$&
$\frac{g_{x}^2}{4 \pi} (f_{x}/g_{x})$&
$m_{x}$ [MeV] & $\Lambda_{x}$ [MeV] & $n_{x}$ \\
\hline
$\pi$  & 14.4 & 138.03 & 1700 & 2 \\
$\rho$ & 0.9 (6.1) & 769 & 1500 & 1 \\
$\omega$ & 20.0 (0) & 782.6 & 1500 & 1 \\
$\sigma$ & 9.4080 & 575 & 2600 & 1 \\
$\delta$ & 0.3912 & 983 & 1500 & 1 \\
$\eta$ & 5.0 & 548.8 & 1500 & 1 \\
\end{tabular}
\end{center}
 \label{kap3_tab_cutoff}
\end{table}

\begin{table}
\caption{$NN$  and  $N\Delta$ partial waves ($^{2s+1}l_j$) with total 
angular momentum $j \leq 3$, parity $\pi$ and isospin $t$.}
\begin{center}
\begin{tabular}{ccccc} 
 $j$ & $\pi$ &t&  $NN$ & $N\Delta$ \\ 
\hline 
 0 & + & 1& $^1S_0$ & $^5D_0$ \\
 0 & $-$  &1& $^3P_0$ & $^3P_0$    \\ \hline
 1 & $-$ & 1& $^3P_1$ & $^3P_1$, $^5P_1$, $^5F_1$ \\ 
 1 & + & 0 & $^3S_1$, $^3D_1$ & \\ 
 1 & $-$ & 0 & $^1P_1$ & \\ \hline
 2 & + &  1& $^1D_2$ & $^5S_2$, $^3D_2$, $^5D_2$, $^5G_2$  \\
 2 & $-$ &1  &$^3P_2$, $^3F_2$ & $^3P_2$, $^5P_2$, $^3F_2$, $^5F_2$ \\
 2 & + & 0 & $^3D_2$ & \\ \hline 
 3 & $-$ & 1 &$^3F_3$ & $^5P_3$, $^3F_3$, $^5F_3$, $^5H_3$ \\
 3 & + & 0 & $^3D_3$, $^3G_3$ & \\
 3 & $-$ & 0 & $^1F_3$  & \\ 
\end{tabular}
\end{center}
\label{kap6_partial}
\end{table}

\begin{table}
\caption{ Parameter values for $V^0_{{\bar N}\Delta}$ and 
$V^0_{\Delta \Delta}$ in the retarded approach.}
\begin{center}
\begin{tabular}{cccccccc}
CC(type) &
$\frac{\left({\bar f}^0_{\Delta N\pi}\right)^2}{4\pi}$ &
${\bar \Lambda}_{\Delta N\pi}$ [MeV]&
${\bar n}_{\Delta\pi}$ &
$\frac{\left({\bar f}^0_{\Delta N\rho}\right)^2}{4\pi}$ &
${\bar \Lambda}_{\Delta N\rho}$ [MeV]&
${\bar n}_{\Delta\rho}$ &
$\alpha_{\Delta N\rho}$ \\
\hline
$(\mbox{ret},\pi,\rho,-1)$ &
0.224 &
1200 &
1 &
20.45 &
1340 &
2 &
$-1$ \\
$(\mbox{ret},\pi,\rho,0)$ &
0.224 &
1200 &
1 &
20.45 &
1420 &
2 &
0 \\
$(\mbox{ret},\pi,\rho,1)$ &
0.224 &
1200 &
1 &
20.45 &
1560 &
2 &
1 \\
\end{tabular}
\end{center}
 \label{kap6_tab1}
\end{table}

\begin{table}
\caption{Parameter values in the static approach. Note that the models
differ in the parametrization of the $\pi N \Delta$ vertex 
$V^0_{\Delta \pi}$ (see the discussion in the text).}
\begin{center}
\begin{tabular}{cccccccc}
CC(type) &
$\frac{\left(f^0_{\pi}\right)^2}{4\pi}$ &
${\Lambda}_{\pi}$ [MeV]&
${n}_{\pi}$ &
$\frac{\left(f^0_{\rho}\right)^2}{4\pi} (f^0_{\rho}/g^0_{\rho})$ &
${\Lambda}_{\rho}$ [MeV] &
${n}_{\rho}$ \\
\hline
$(\mbox{stat1},\pi)$ &
0.08 &
700 &
1 &
0 (0) &
  &\\
$(\mbox{stat2},\pi)$ &
0.08 &
680 &
1 &
0 (0)  &
  &\\
$(\mbox{stat},\pi,\rho,0)$ &
0.0778 &
1300 &
1 &
0.84 (6.1) &
1400 &
1 \\
\hline
\hline
CC(type) &
$\frac{\left({\bar f}^0_{\Delta N\pi}\right)^2}{4\pi}$ &
${\bar \Lambda}_{\Delta N\pi}$ [MeV]&
${\bar n}_{\Delta\pi}$ &
$\frac{\left({\bar f}^0_{\Delta N\rho}\right)^2}{4\pi}$ &
${\bar \Lambda}_{\Delta N\rho}$ [MeV]&
${\bar n}_{\Delta\rho}$ &
$\alpha_{\Delta N\rho}$ \\
\hline
$(\mbox{stat1},\pi)$ &
0.35 &
700 &
1 &
  &
  &
  &\\
$(\mbox{stat2},\pi)$ &
0.35 &
680 &
1 &
  &
  &
  &\\
$(\mbox{stat},\pi,\rho,0)$ &
0.224 &
1200 &
1 &
20.45 &
1600 &
2 &
0 \\
\end{tabular}
\end{center}
 \label{kap6_tab2}
\end{table}

\begin{figure}[hp]
\centerline{\psfig{figure=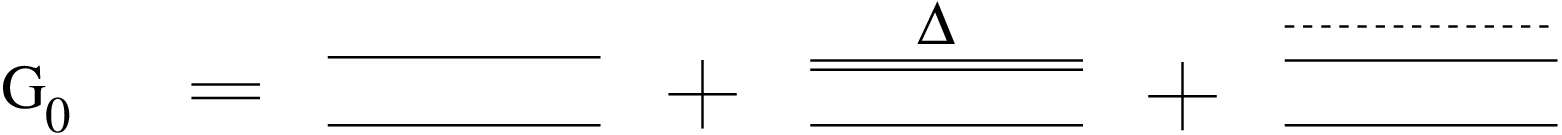,width=12cm,angle=0}}
\vspace{0.5cm}
\caption{Diagrammatic representation of the free propagator $G_0(z)$ with 
the three contributions in ${\cal H}^{[2]}_{\bar N }$, 
${\cal H}^{[2]}_{\Delta}$, and 
${\cal H}^{[2]}_{X}$.}
\label{fig_G_0}
\end{figure}

\newpage

\newpage

\end{document}